\documentclass[aps,prc,twocolumn,superscriptaddress,showpacs,preprintnumbers,amsmath,amssymb,floatfix]{revtex4}

\usepackage{graphicx}
\usepackage{dcolumn}
\usepackage{bm}
\usepackage{longtable}
\usepackage{color}

\begin{document}

\preprint{MAX-lab $\bm^{4}$He$\bm(\gamma,n)$ Summary Article Version 6.0}

\title{A measurement of the $^{4}$He$(\gamma,n)$ reaction from 23 $\bm<$ E$\bm_{\gamma}$ $\bm<$ 70 MeV}

\author{B. Nilsson}
\affiliation{Department of Physics, University of Lund, SE-221 00 Lund, Sweden}
\author{J. -O. Adler}
\affiliation{Department of Physics, University of Lund, SE-221 00 Lund, Sweden}
\author{B. -E. Andersson}
\altaffiliation{Present address:  Furulundsskolan, SE-29480 S\"{o}lvesborg, Sweden}
\affiliation{Department of Physics, University of Lund, SE-221 00 Lund, Sweden}
\author{J. R. M. Annand}
\affiliation{Department of Physics and Astronomy, University of Glasgow, G12 8QQ Glasgow, UK}
\author{I. Akkurt}
\altaffiliation{Present address:  Department of Physics, S\"{u}leyman Demirel University, Fen-Edebiyat Faculty, 32 260 Isparta, Turkey.}
\affiliation{Department of Physics and Astronomy, University of Glasgow, G12 8QQ Glasgow, UK}
\author{M. J. Boland}
\altaffiliation{Present address:  Australian Synchrotron, Clayton, Victoria 3168, Australia}
\affiliation{Department of Physics, University of Lund, SE-221 00 Lund, Sweden}
\author{G. I. Crawford}
\affiliation{Department of Physics and Astronomy, University of Glasgow, G12 8QQ Glasgow, UK}
\author{K. G. Fissum}
\altaffiliation{Corresponding author; \texttt{kevin.fissum@nuclear.lu.se}}
\affiliation{Department of Physics, University of Lund, SE-221 00 Lund, Sweden}
\author{K. Hansen}
\altaffiliation{Present address:  MAX-lab, University of Lund, SE-221 00 Lund, Sweden}
\affiliation{Department of Physics, University of Lund, SE-221 00 Lund, Sweden}
\author{P. D. Harty}
\altaffiliation{Present address:  \texttt{Peter.Harty@dhs.vic.gov.au}}
\affiliation{Department of Physics and Astronomy, University of Glasgow, G12 8QQ Glasgow, UK}
\author{D. G. Ireland}
\affiliation{Department of Physics and Astronomy, University of Glasgow, G12 8QQ Glasgow, UK}
\author{L. Isaksson}
\altaffiliation{Present address:  MAX-lab, University of Lund, SE-221 00 Lund, Sweden}
\affiliation{Department of Physics, University of Lund, SE-221 00 Lund, Sweden}
\author{M. Karlsson}
\affiliation{Department of Physics, University of Lund, SE-221 00 Lund, Sweden}
\author{M. Lundin}
\altaffiliation{Present address:  MAX-lab, University of Lund, SE-221 00 Lund, Sweden}
\affiliation{Department of Physics, University of Lund, SE-221 00 Lund, Sweden}
\author{J. C. McGeorge}
\affiliation{Department of Physics and Astronomy, University of Glasgow, G12 8QQ Glasgow, UK}
\author{G. J. Miller}
\altaffiliation{Present address: NEL Oil and Gas Services, TUV NEL Ltd., G75 0QU East Kilbride, UK.}
\affiliation{Department of Physics and Astronomy, University of Glasgow, G12 8QQ Glasgow, UK}
\author{H. Ruijter}
\altaffiliation{Present address:  Sony Ericsson Mobile Communications AB, SE-221 88 Lund, Sweden}
\affiliation{Department of Physics, University of Lund, SE-221 00 Lund, Sweden}
\author{A. Sandell}
\altaffiliation{Present address:  Lund University Hospital,  SE-22185 Lund, Sweden }
\affiliation{Department of Physics, University of Lund, SE-221 00 Lund, Sweden}
\author{B. Schr\"{o}der}
\affiliation{Department of Physics, University of Lund, SE-221 00 Lund, Sweden}
\author{D. A. Sims}
\altaffiliation{Present address:  \texttt{David.Sims@tenix.com}}
\affiliation{Department of Physics, University of Lund, SE-221 00 Lund, Sweden}
\author{D. Watts}
\altaffiliation{Present address:  School of Physics, University of Edinburgh, EH9 3JZ Edinburgh, UK}
\affiliation{Department of Physics and Astronomy, University of Glasgow, G12 8QQ Glasgow, UK}

\collaboration{The MAX-lab Nuclear Physics Working Group}
\noaffiliation

\date{\today}

\begin{abstract}

A comprehensive set of $^{4}$He$(\gamma,n)$ absolute cross-section measurements
has been performed at MAX-lab in Lund, Sweden.  Tagged photons from 23 $<$ 
$E_{\gamma}$ $<$ 70 MeV were directed toward a liquid $^{4}$He target, and 
neutrons were identified using pulse-shape discrimination and the 
time-of-flight technique in two liquid-scintillator detector arrays.
Seven-point angular distributions have been measured for fourteen photon 
energies.  The results have been subjected to complementary 
Transition-coefficient and Legendre-coefficient analyses.  The results are also
compared to experimental data measured at comparable photon energies as well as
Recoil-Corrected Continuum Shell Model, Resonating Group Method, and
Effective Interaction Hyperspherical-Harmonic Expansion calculations.  For 
photon energies below 29 MeV, the angle-integrated data are significantly 
larger than the values recommended by Calarco, Berman, and Donnelly in 1983.

\end{abstract}


\maketitle

\section{\label{section:intro}Introduction}

Over the past 50 years, a very large body of experimental work has been 
performed in order to understand the near-threshold photodisintegration of 
$^{4}$He.  In 1983, Calarco, Berman, and Donnelly (CBD) assessed all of the 
available experimental data in a benchmark review article \cite{calarco83} and 
made a recommendation as to the value of the $^{4}$He$(\gamma,n)$ 
photodisintegration cross section from threshold up to $E_{\gamma}$ $\sim$ 50
MeV.  Since then, most of the experimental effort has been directed towards
measuring either the ratio of the photoproton-to-photoneutron cross sections or
simply the photoproton channel.  Up-to-date reviews of all available data are 
made in Refs. \cite{quaglioni04,shima05}.  In contrast to the $(\gamma,p)$ 
situation, only three near-threshold measurements of the photoneutron channel 
have to our knowledge been published \cite{komar93,shima01,shima05}.

In this Paper, we present a comprehensive new data set for the 
$^{4}$He$(\gamma,n)$ reaction near threshold which has been obtained using 
tagged photons with energies from 23 $<$ $E_{\gamma}$ $<$ 70 MeV.  We compare 
our data with the CBD evaluation as well as the post-CBD data.  We also report 
the results of Transition-coefficient and Legendre-coefficient analyses of our 
data, and compare them to Recoil-Corrected Continuum Shell Model (RCCSM) 
calculations \cite{halderson81,halderson04}, a Resonating Group Method (RGM) 
calculation \cite{wachter88}, and an Effective Interaction Hyperspherical
Harmonic (EIHH) Expansion calculation \cite{quaglioni04}.  A detailed 
description of the experiment is presented in Ref. \cite{nilsson03}, and 
preliminary findings have been sketched in Refs. \cite{sims98,nilsson05}.

\section{\label{section:expt}Experiment}

\begin{figure*}
\begin{center}
\resizebox{0.85\textwidth}{!}{\includegraphics{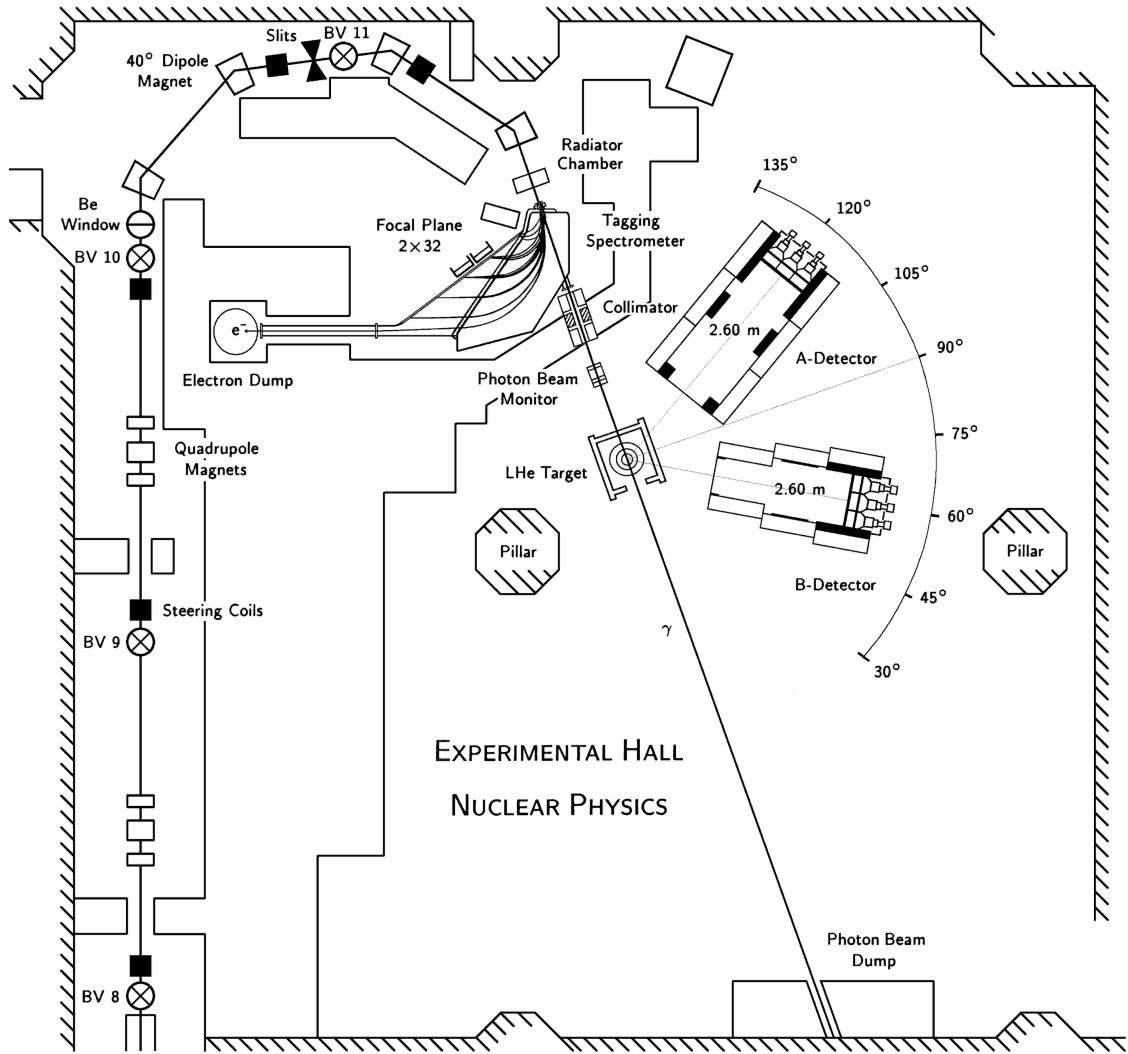}}
\end{center}
\caption{\label{figure:exphall}
The photonuclear hall at MAX-lab at the time of the experiment.  The electron 
beam passed through a 0.1\% radiation-length Al radiator generating 
bremsstrahlung.  Non-interacting electrons were dumped.  Recoil electrons were 
momentum-analyzed using a magnetic tagging spectrometer equipped with a 
64-counter focal-plane scintillator array.  The resulting tagged-photon beam 
was collimated before it struck the storage-cell liquid-$^{4}$He target (see 
Figure \ref{figure:target}).  Knocked-out neutrons were detected in two 
movable, large solid-angle, liquid-scintillator arrays (see Figure 
\ref{figure:ndetector}).  See text for further details.}
\end{figure*}

The experiment was performed at the tagged-photon facility \cite{adler97} 
located at MAX-lab \cite{maxlab}, in Lund, Sweden.  A pulse-stretched electron 
beam with an energy of $\sim$93 MeV, a current of $\sim$30 nA, and a duty 
factor of $\sim$75\% was used to produce quasi-monoenergetic photons via the 
bremsstrahlung-tagging technique \cite{adler90}.  A diagram of the experimental 
layout is shown in Figure \ref{figure:exphall}.

\subsection{\label{subsection:beam}Photon beam}

A 0.1\% radiation-length aluminum radiator was used to convert the incident 
electron beam into bremsstrahlung.  Non-radiating electrons were dumped into 
a Faraday cup which recorded the electron-beam current.  This cup was 
surrounded by borated water, lead, and concrete shielding.  Post-bremsstrahlung
electrons were momentum-analyzed using a magnetic spectrometer equipped with a 
64-counter focal-plane scintillator array.  The photon-energy resolution of 
$\sim$300 keV resulted almost entirely from the 10 mm width of a single 
focal-plane counter.  The scintillators were mounted in two 32-counter modules,
and photon-energy ranges were selected by sliding the array to the appropriate 
position along the focal plane of the spectrometer.  The average single-counter
rate during these measurements was 0.5 MHz.

The size of the photon beam was defined by a tapered tungsten-alloy primary 
collimator.  The primary collimator was followed by a dipole magnet and a 
post-collimator which were used to ``scrub" any charged particles produced in 
the primary collimator.  The photon-beam intensity was monitored continuously 
using a crude pair spectrometer which consisted of an array of three 0.5 mm 
thick plastic scintillators.  The position of the photon beam at the target
location was determined by irradiating Polaroid film after every adjustment of 
the electron beam.  The beam spot was typically 2 cm in diameter at this 
position.

The tagging efficiency \cite{adler97} is the ratio of the number of tagged 
photons which struck the target to the number of recoil electrons which were 
registered by the associated focal-plane counter.  It was measured both 
absolutely (using a 100\% efficient lead/scintillating-fiber photon detector) 
and relatively (using the pair-spectrometer beam monitor) during the 
experiment.  The absolute measurements required a very low intensity photon 
beam to avoid pileup in the photon detector, and were performed periodically 
throughout the experiment.  The relative measurement was made continuously.  
Tagging efficiency was typically $\sim$25\%.

\subsection{\label{subsection:target}Cryogenic target}

\begin{figure}
\begin{center} \resizebox{0.44\textwidth}{!}{\includegraphics{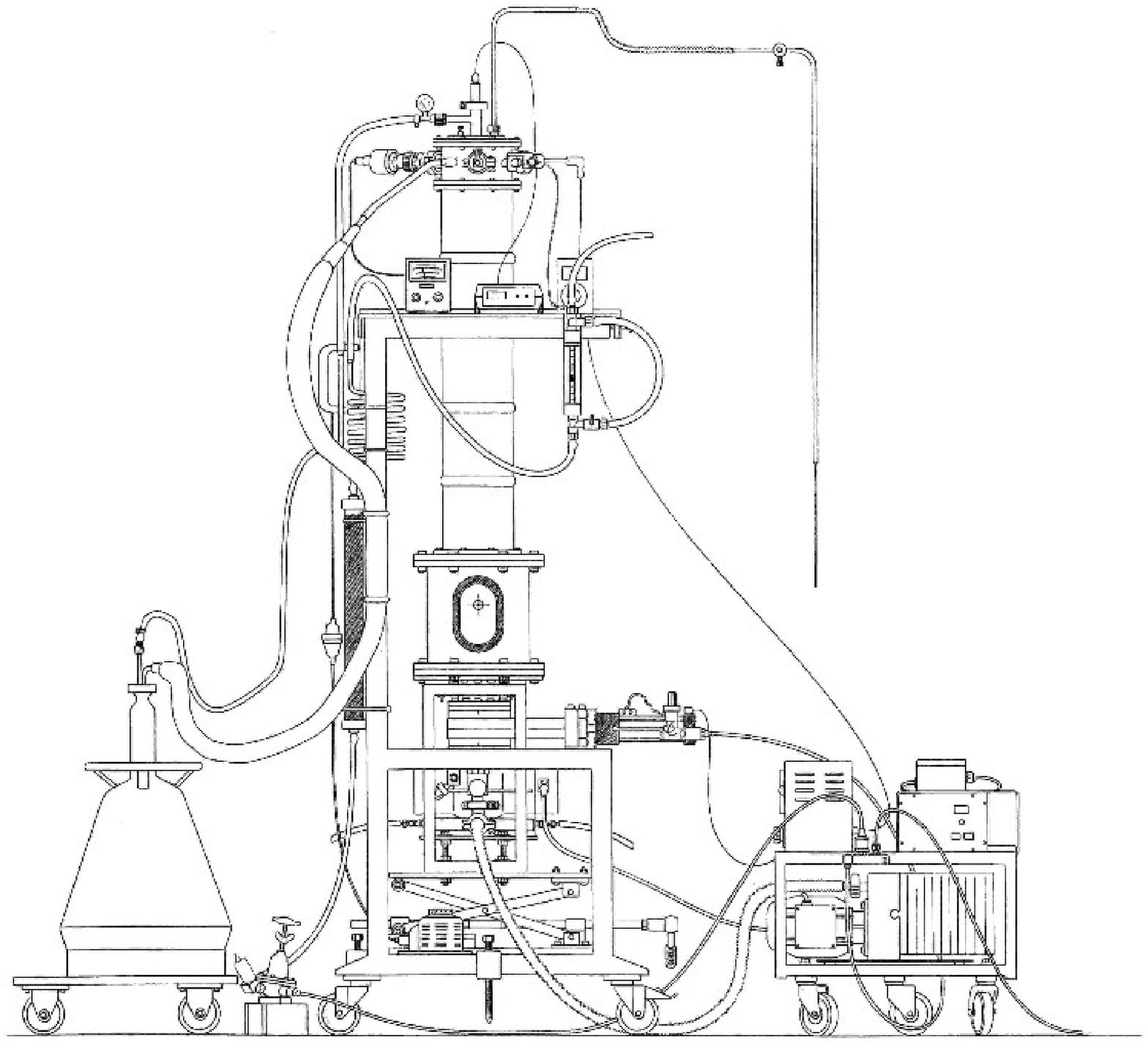}}
\end{center}
\begin{center} \resizebox{0.44\textwidth}{!}{\includegraphics{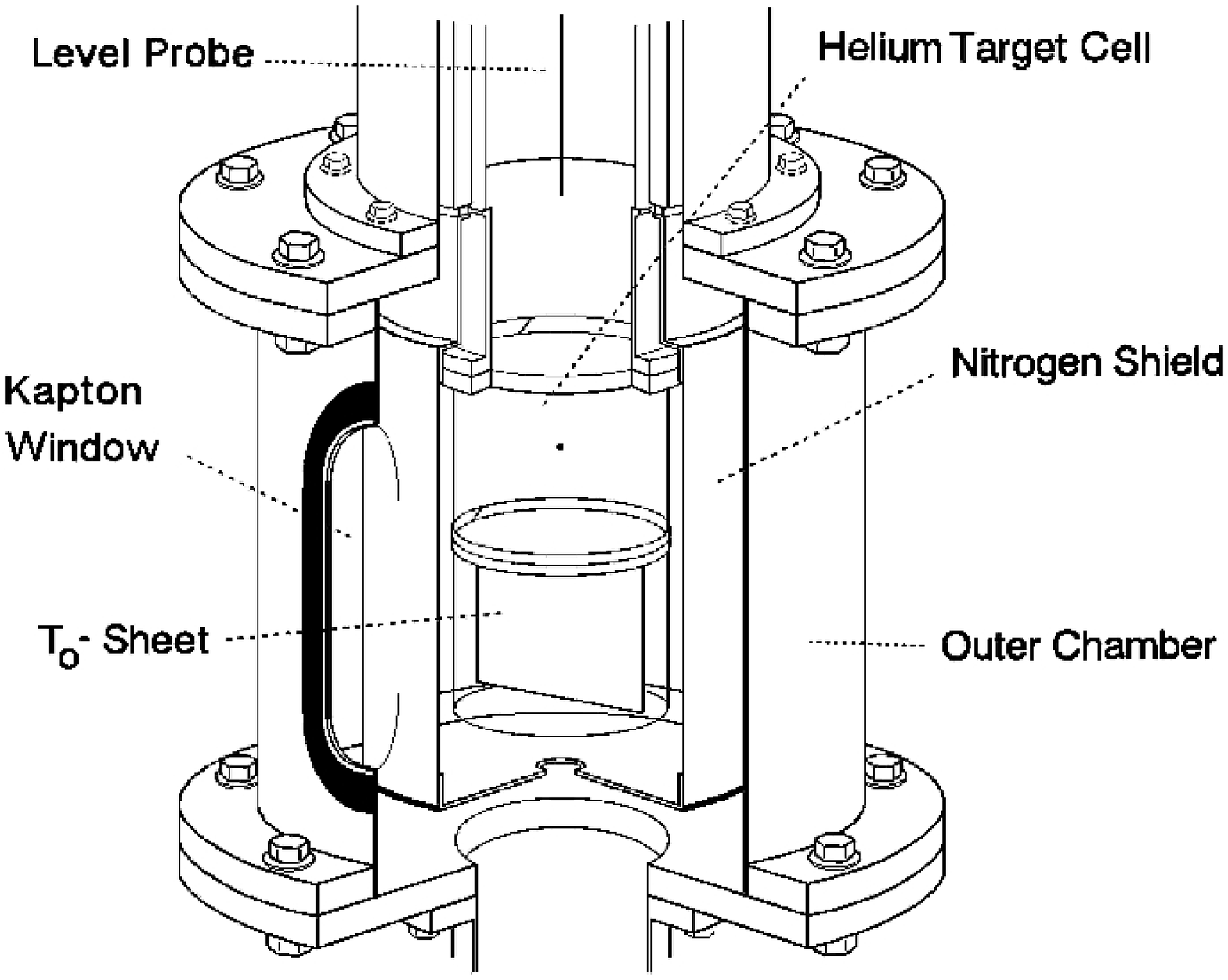}}
\end{center}
\caption{\label{figure:target}
The cryogenic target.  The top panel shows the entire cryogenic target system 
(including the vacuum pumps and the filling, monitoring, and support 
subsystems), while the bottom panel shows just the target chamber, in which two 
target cells were stacked vertically on top of one another.  The upper cell
contained the liquid $^{4}$He and was also used for the empty-target 
measurements, while the lower cell contained the 1 mm thick steel sheet used 
for TOF calibration of the neutron detectors.  The cells could be moved in and 
out of the photon beam in the laboratory vertical direction using a ladder 
mechanism.  See text for further details.
}
\end{figure}

Liquid $^{4}$He was provided by a 6 liter, top-loaded, storage-cell 
cryostat \cite{nilsson03}, which was refilled on a $\sim$24 hour basis.
The cylindrical target cell (see Figure \ref{figure:target}) was made of 80
$\mu$m thick Kapton foil, and had a diameter of 90 mm and a height of 75 mm.
It was mounted with the cylinder axis perpendicular to the direction of the
photon beam and the reaction plane.  The level of the liquid $^{4}$He in the 
cryostat was continuously monitored throughout the experiment by measuring the 
resistance of a superconducting NbTi probe.  Radiative heating of the target 
cell was reduced using a heat shield of three layers of 30 $\mu$m thick Al foil
and about ten layers of the super-insulation NRC-2, all maintained at 
liquid-N$_{2}$ temperature.  The assembly sat in a 2 mm thick stainless-steel
vacuum chamber with 125 $\mu$m thick Kapton entrance and exit windows.  Vacuum 
was maintained using a water-cooled double-flow turbomolecular pump.  The 
vacuum pressure in the target dewar zones was about 2 $\times$ 10$^{-7}$ mbar, 
which was well below the critical accomodation pressure \cite{nilsson03} of 4.4
$\times$ 10$^{-6}$ mbar.  Density fluctuations in the liquid $^{4}$He were 
inferred from the rate of evaporation \cite{tate83}.  This was monitored 
continuously using both the superconducting level probe and a gas-flow meter 
which measured the rate of outgassing of the evaporating $^{4}$He.  The 
flow-rate fluctuations were negligible \cite{nilsson06b}, so that the density 
of the liquid helium employed in the experiment was 125.20 mg/cm$^{3}$ $\pm$ 
0.01\%.

Empty target-cell measurements were used to determine the non-$^{4}$He 
background, which was negligible.  A 1 mm thick steel sheet mounted below the 
cell on the movable target ladder was used to convert photons to relativistic
$e^{+}e^{-}$ pairs for TOF calibration of the neutron detectors (see below).

\subsection{\label{subsection:neutron_detectors}Neutron detectors}

\subsubsection{\label{subsubsection:detectors_general}General properties}

Neutrons were detected in two large solid-angle neutron detectors
\cite{annand97}.  Each detector consisted of a 3 $\times$ 3 array of 9 
rectangular cells with internal dimensions 20 cm $\times$ 20 cm $\times$ 10 cm 
(deep) filled with the liquid scintillator NE213A.  Each cell was instrumented
with a 5" photomultiplier (model 9823) from Thorn EMI.  The arrays were mounted 
on movable platforms (30 $\deg$ $<$ $\theta_{\rm neutron}$ $<$ 135 $\deg$) and 
encased in Pb, steel, and borated-wax shielding.  Plastic scintillators which 
were 65 cm $\times$ 65 cm $\times$ 2 cm (thick) were placed in front of the 
arrays and used to identify and veto incident charged particles.  Each of the
veto scintillator paddles was instrumented with two 2" photomultipliers (model 
XP2262B) from Philips.  Since only moderate energy resolution was necessary to 
identify the two-body photodisintegration of $^{4}$He unambiguously, the 
detectors were placed $\sim$2.6 m (a relatively short flight path) from the 
target.  The resulting nominal geometrical solid angle subtended by a single 
cell within the array was $\sim$6 msr.

\begin{figure}
\begin{center} \resizebox{0.47\textwidth}{!}{\includegraphics{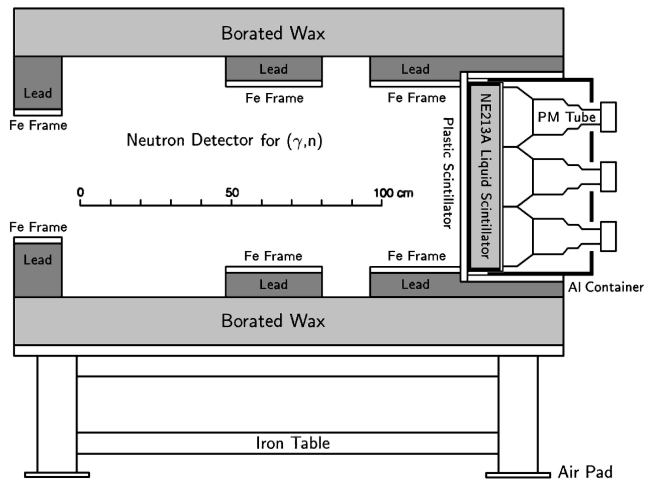}}
\end{center}
\caption{\label{figure:ndetector}
A sideview of one of the neutron detectors.  The 9-cell detector array together
with the plastic-scintillator veto and the associated shielding sat on table
which could be moved using airpads.  See text for further details.}
\end{figure}

\subsubsection{\label{subsubsection:detectors_psd}Pulse-shape discrimination (PSD)}

The PSD technique was employed in this experiment to distinguish between
neutrons and photons.  This technique relies on the fact that the shape of the
scintillation pulse in the NE213A scintillator is dramatically different for
neutrons and photons.  The scintillation pulse from NE213A has both fast ($<$5 
ns) and slow ($\sim$500 ns) decay components whose relative intensities depend 
upon the density of the ionization along the track of the interacting particle. 
Highly ionizing, non-relativistic protons from neutron-induced reactions in 
the liquid scintillator have an enhanced slow-decay intensity compared to
electrons resulting from photon conversion.  The total charges resulting from 
pulse-integration periods of 25 and 500 ns were compared using purpose-built 
hardware \cite{annand87} which provided both a comparison (difference) analog 
output (the ``pulse-shape" or ``PS" signal) and a logic output which signaled 
that the comparison voltage-level threshold had been crossed.  The latter was 
used in the hardware event trigger.

\subsubsection{\label{subsubsection:detectors_tof}Time-of-flight (TOF) technique}

\begin{figure}
\begin{center}
\resizebox{0.47\textwidth}{!}{\includegraphics{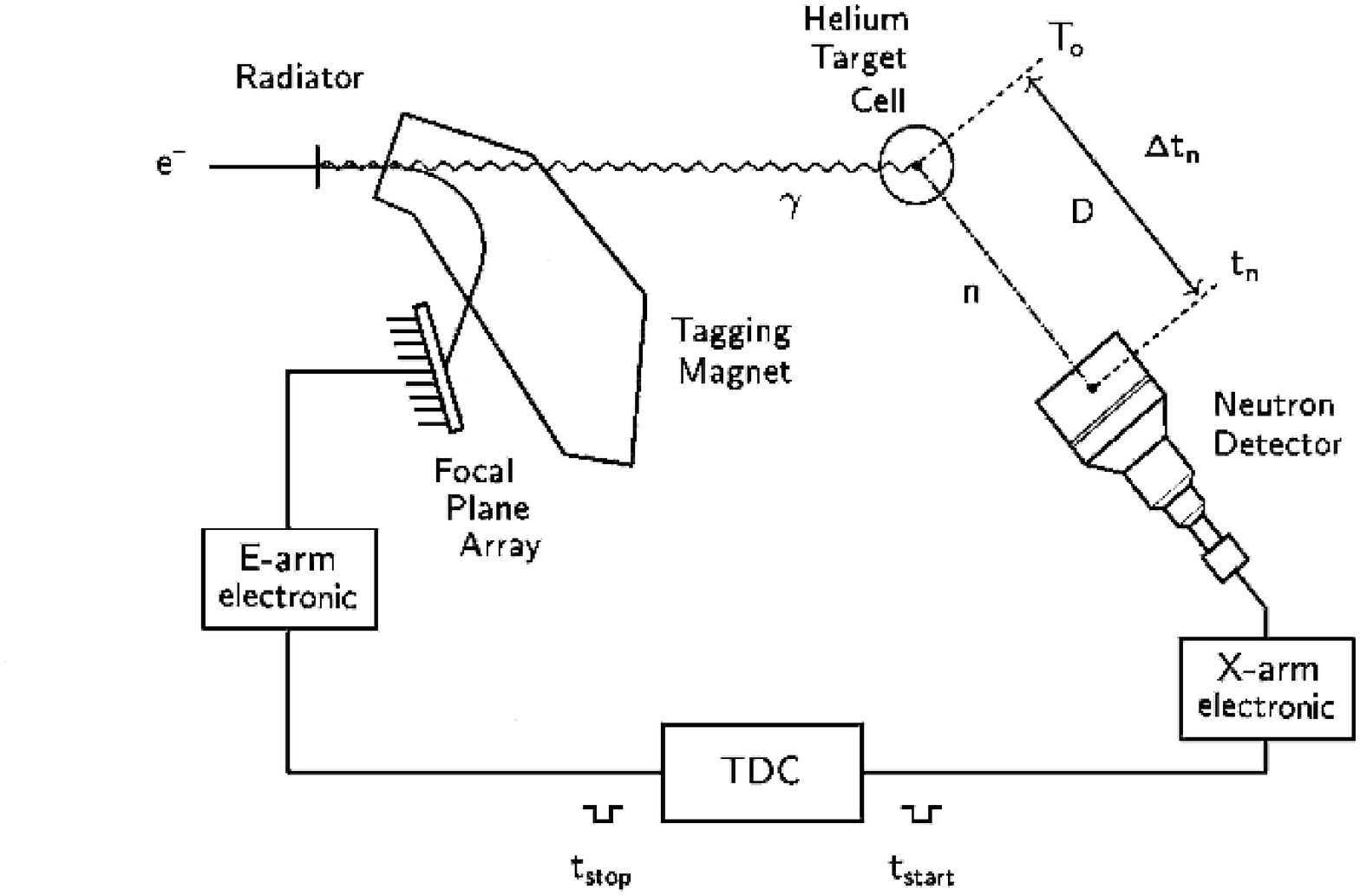}}
\end{center}
\begin{center}
\resizebox{0.44\textwidth}{!}{\includegraphics{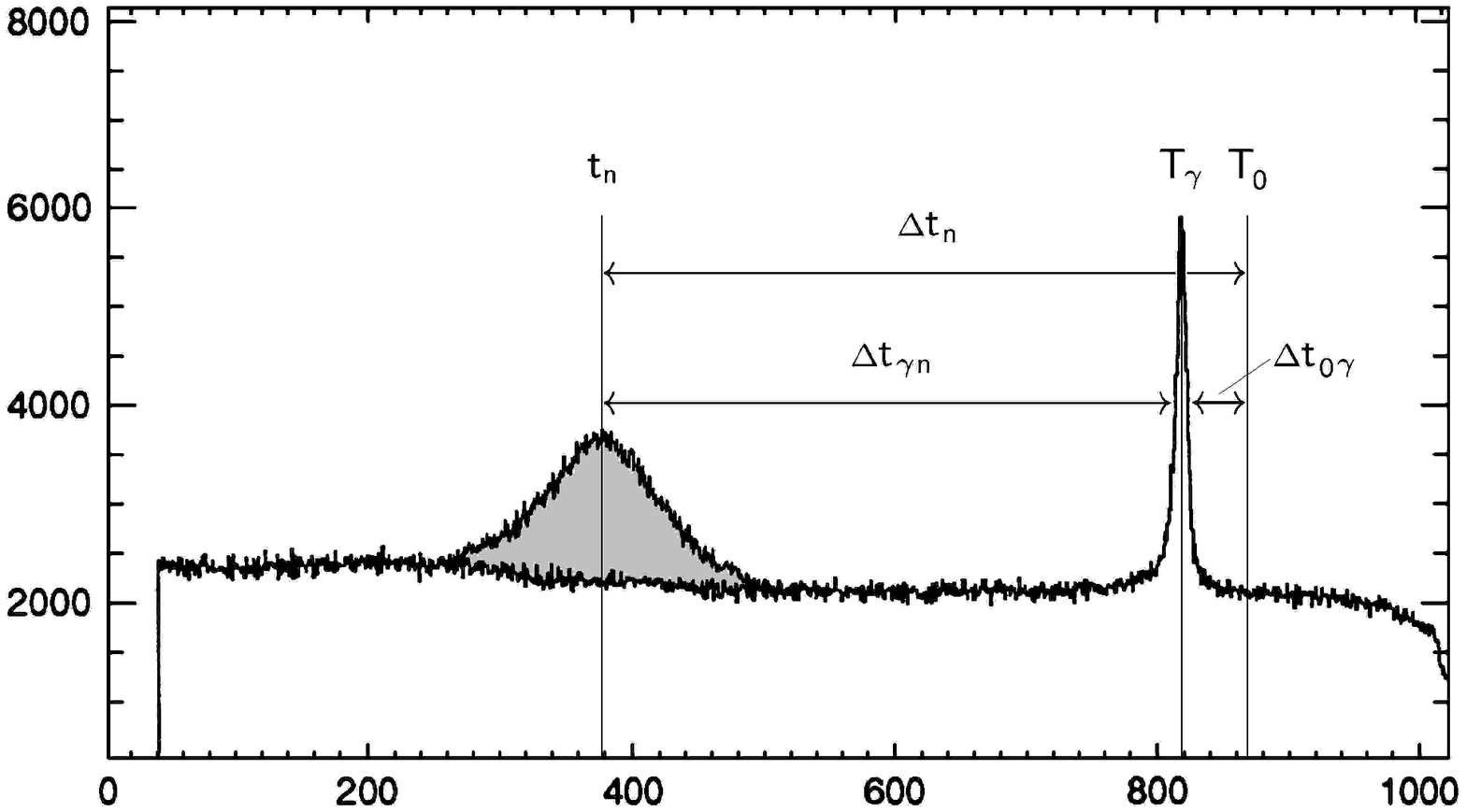}}
\vspace*{-2.75cm}\\ \hspace*{-8.5cm}\rotatebox{90} {\textsf{Events}}\\
\vspace*{1.8cm} {\hspace*{0.70cm}\textsf{TOF  (TDC channels)}}\\
\end{center}
\caption{\label{figure:tof_technique}
An illustration of the TOF technique.  In the top panel, an overview of the
TOF measurement is presented.  A tagged photon knocks a neutron out of the
$^{4}$He target which is then detected in the neutron-detector cell.  This 
signal starts a TDC.  The corresponding recoil electron is detected in the 
tagger focal plane, which results in a stop signal for the TDC.  As a result, 
the TOF (and thus the neutron energy) is measured.  In the bottom panel, a 
corresponding TOF spectrum is shown.  Quasimonoenergetic neutrons are displayed
as a gray bump superimposed on a random background.  See text for further
details.
}
\end{figure}

The TOF technique was employed to determine the neutron energy.  The principles
of this technique are demonstrated in Figure \ref{figure:tof_technique}.  In 
the top panel, the tagged photon knocks a neutron out of the liquid-$^{4}$He 
target cell at time $T_{0}$.  This neutron is subsequently detected in a 
neutron-detector cell located a distance $D$ from the target at time $t_{n}$.  
The neutron-detector signal is used to start a time-to-digital converter (TDC),
which is then stopped by the signal from the recoil electron striking the 
tagger focal plane.  The TDC thus measures the neutron TOF ($\Delta t_{n}$) 
from which the neutron kinetic energy $T_{n}$ is obtained.  In the bottom 
panel, a typical TOF spectrum is shown for neutrons of approximate energies 2.5
$<$ $T_{n}$ $<$ 6.5 MeV, which are displayed as the shaded bump.  The width of 
this bump results from the neutron-energy range, the flight-path uncertainty, 
and any electronic pick-off uncertainty.  The ``photon" calibration peak 
labeled $T_{\gamma}$ lying between the neutrons and the $T_{0}$ position is 
much narrower, reflecting the lack of any velocity-dependent broadening.  
Often referred to as the ``gamma flash'', it results in principle from a TOF 
measurement of the photons originating in the target which made it past the 
hardware PSD (see Section \ref{subsection:daq}).  This ``light-speed" peak is a
useful calibration point from which $T_{0}$ can be calculated.  For measurement 
purposes, it was enhanced by switching the plastic veto detectors in front
of the neutron arrays to coincidence mode so that relativistic electrons 
produced in the target were also registerd.  A full-width-at-half-maximum TOF 
neutron-energy resolution of better than 2 MeV for all photon energies was 
obtained in this experiment.  Further, as a result of the two-body kinematics, 
the measured energy of the neutrons provided a cross check upon the 
tagged-photon energy.

\subsection{\label{subsection:daq}Electronics and data acquisition}

\begin{figure}
\begin{center}
\resizebox{0.47\textwidth}{!}{\includegraphics{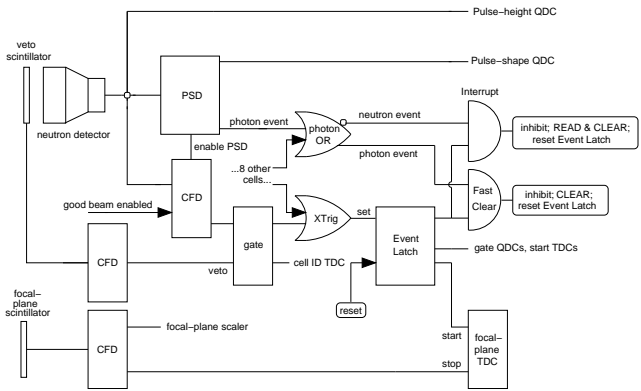}}
\end{center}
\caption{\label{figure:electronics}
An overview of the experiment electronics.  See text for further details.
}
\end{figure}

An overview of the experiment electronics is presented in 
Figure \ref{figure:electronics}.  The analog signals from the liquid 
scintillators were symmetrically divided three ways and passed to a 
charge-to-digital converter (QDC), a PSD module, and a constant-fraction 
discriminator (CFD).  The logical OR of the CFD signals, in anti-coincidence 
with the equivalent veto-detector output, formed a primary event trigger which 
was used to start the analog-to-digital circuitry.  

The PSD modules were used to identify photons.  When photons were identified,
the logical outputs from the PSDs were used to abort the event processing and 
fast clear the QDCs and TDCs.  When photons were not identified, the event
was read out.  The PSD thresholds were set conservatively to avoid the 
rejection of any neutrons, and as a result, a small fraction of the otherwise 
overwhelming number of background photon events ``leaked" into the data set 
(see Figure \ref{figure:scatterplot}).

No hardware coincidence was made between the neutron TOF spectrometers and the 
tagger focal-plane array, but coincidences were recorded in the TDCs attached 
to the 64 focal-plane counters.  The neutron detectors made the common start.  
Focal-plane rates were recorded in the 64 free-running scalers also attached to
the focal-plane counters.  As the scalers were not inhibited during the 
data-acquisition dead time, proper normalization of the neutron yield required 
a livetime-efficiency correction.  This correction was simply the ratio of the 
number of processed events (given by the sum of the scalers counting the number
of read \& clears and fast clears) to the number of potential triggers (which
came from a free-running scaler counting the OR of the CFD signals).  The 
system livetime was also estimated using two other methods which are described 
in detail in Ref. \cite{nilsson03}.  The first method considered the interrupt 
rates of read \& clears and fast clears together with their processing times 
and the duty factor of the beam.  The second method considered the outputs of 
free-running and inhibited oscillators together with the duty factor of the 
beam.  All three methods yielded the same results.

\begin{figure}
\begin{center}
\resizebox{0.47\textwidth}{!}{\includegraphics{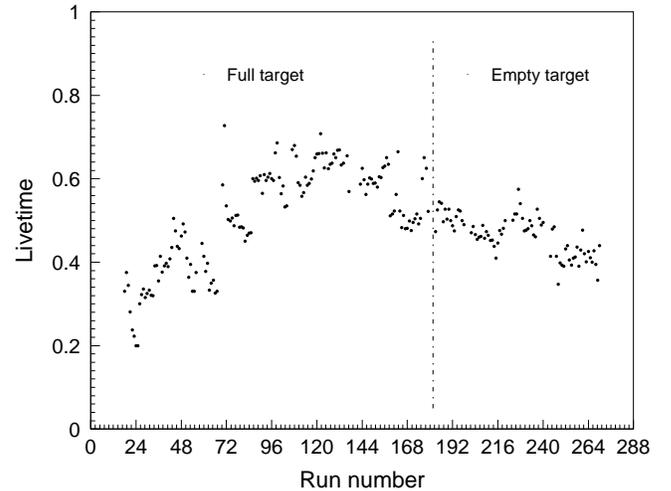}}
\end{center}
\caption{\label{figure:livetime}
The livetime efficiency correction as a function of run number.  Variations in
the correction are due to variations in the event rate due to different
electron-beam intensities and different neutron-detector angles.  See text for 
further details.
}
\end{figure}

Figure \ref{figure:livetime} shows the distribution of livetime over the 
duration of the experiment, with each point in the Figure representing a single
1-hour run.  As previously mentioned, the hardware PSD was purposely set rather
``loose" so that no neutrons could accidentally be rejected.  As a result, 
background photon events which made it into the data stream contributed
overwhelmingly to the livetime of the system.  As the neutron detectors were 
sensitive to even the smallest fluctuations in the electron beam, run-to-run 
variations in the livetime occurred for otherwise identical experiment 
configurations.  Further, the background level was a strong function of angle, 
which also led to variations in the system livetime.  
The mean value was $\sim$50\%.

Data acquisition and storage were handled using an in-house 
toolkit \cite{ruijter95}.  Subsequent offline analysis was performed using the 
program \textsc{acqu} \cite{annand_acqu}.

\section{\label{section:analysis}Analysis}

\subsection{\label{subsection:energy_calibration}Neutron-detector energy calibration}

Gamma rays with energies ranging from 1.2 $<$ $E_{\gamma}$ $<$ 7.1 MeV from the
sources $^{60}$Co and $^{239}$PuBe were used to calibrate the energy deposited 
in the NE213A scintillators by measuring the pulse-height distributions from 
recoiling Compton electrons \cite{annand97}.  The method reported in 
Ref. \cite{knox72} was used to determine the Compton edge position, but other 
prescriptions \cite{flynn64,king84} did not produce a significantly different 
calibration.  The calibration was necessary to determine the neutron-detection 
threshold (in MeV$_{ee}$ or ``MeV electron-equivalent") which in turn was 
necessary to calculate the neutron-detection efficiency (see below).  The 
non-linear response of NE213A to low-energy recoil protons was modeled using 
the empirical expression of Ref. \cite{cecil79}.  In general, the 
neutron-detection efficiency was very sensitive to the neutron pulse-height 
threshold, and thus a precise energy calibration of the NE213A scintillators 
was crucial.  

\subsection{\label{subsection:pid}Particle identification (PID)}

\begin{figure}
\begin{center}
\resizebox{0.42\textwidth}{!}{\includegraphics{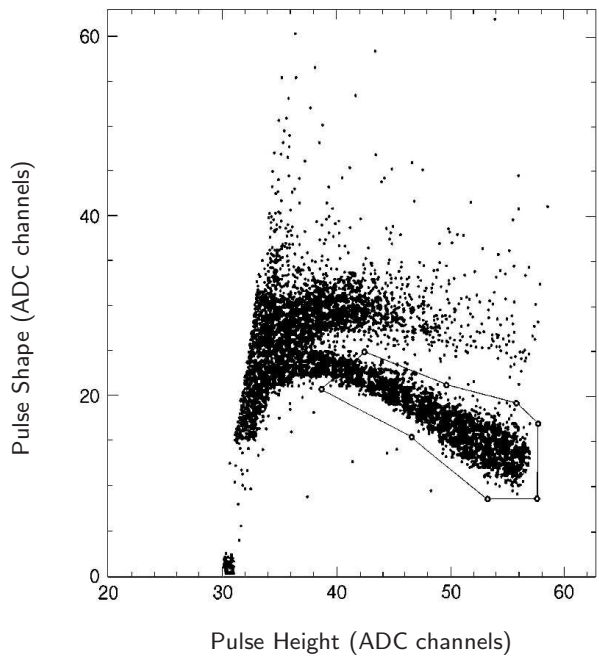}}
\vspace*{-6.2cm}\\
\hspace*{-8.5cm}\rotatebox{90}
{\textsf{Pulse Shape (ADC channels)}}\\
\vspace*{2.2cm}
{\hspace*{0.5cm}\textsf{Pulse Height  (ADC channels)}}\\
\end{center}
\caption{\label{figure:scatterplot}
A typical PID scatterplot for a single neutron-detector cell.  Pulse shape (PS)
was plotted versus pulse height (PH).  Separation between neutrons and random 
photons is clearly demonstrated.  The cut polygon which encircles the photon 
ridge was set very ``loose'' to ensure that no neutron events were discarded.  
Higher PH cuts were applied at a later stage of the data analysis.  See text 
for further details.}
\end{figure}

As previously discussed, a PSD cut made using the trigger-processing 
electronics was used to reject photons at the hardware level, as the background
photon flux was $\sim$10$^{5}$ times greater than the neutron flux.  More 
precise PID was performed offline using the recorded PS amplitude from the PSD 
module, plotted as a function of energy deposited in the NE213A scintillator 
(the ``pulse height" or ``PH" signal).  Figure \ref{figure:scatterplot} shows 
the distinct separation between neutron and photon events which resulted.

\subsection{\label{subsection:background_subtraction}Background removal}

\begin{figure}
\begin{center}
\resizebox{0.47\textwidth}{!}{\includegraphics{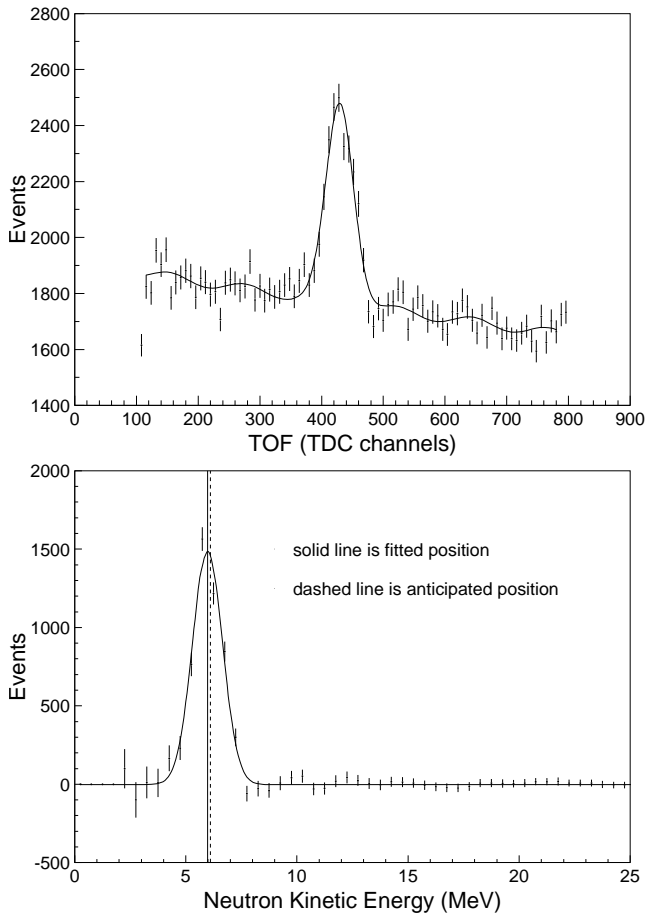}}
\end{center}
\caption{\label{figure:tof}
A typical TOF spectrum measured at $\theta_{\rm LAB}$ $=$ 90 $\deg$ for
$E_{\gamma}$ $=$ 29 MeV.  Also shown in the top panel is the fitted function 
defined in Eq. (\ref{equation:fit_function}).  The prominent peak corresponds 
to neutron events.  The same distribution, this time with the background 
removed and plotted as a function of neutron kinetic energy, is shown in the 
bottom panel.  The vertical solid line is the fitted neutron-peak location 
(6.0 MeV), while the vertical dashed line is the anticipated neutron-peak 
location (6.1 MeV).  See text for further details.}
\end{figure}

After the selection of candidate neutron events, the resulting data set for 
each neutron detector consisted of 64 TOF spectra containing both real 
coincidences with the tagger focal plane and a random background (see the top
panel of Figure \ref{figure:tof}).  The ratio of the number of prompt neutrons 
to random background was a strong function of photon energy, ranging from 
better than 1-to-1 at $E_{\gamma}$ $=$ 68.5 MeV to 1-to-10 at $E_{\gamma}$ $=$ 
24.6 MeV.  

The removal of this random background proved to be a challenging exercise due 
to a periodic ripple in the time structure of the photon beam resulting from 
microstructure in the electron beam extracted from the pulse-stretcher 
ring \cite{hoorebeke93}.  This ripple may be clearly seen in the TOF spectrum
shown in the top panel of Figure \ref{figure:tof}.  Thus, the TOF spectra were 
fitted with a function of the form

\begin{equation}
\label{equation:fit_function}
Y(t) =
A \cdot \exp~(B \cdot t) + C \cdot \sin(D \cdot t) + E \cdot \exp-(\frac{t-F}{\sqrt{2}~G})^2
\end{equation}

\noindent
to determine the random yield.  A prompt Gaussian peak (coefficients E, F, 
and G) was superimposed upon a background which contained a sinusoidal term 
(coefficients C and D) to describe the ripple and an exponential term 
(coefficients A and B) which accounted for deadtime in the detector electronics 
and the single-hit TDCs which instrumented the tagger focal plane.  

The bottom panel of Figure \ref{figure:tof} shows the same distribution as in
the top panel, but this time after the background had been removed and now
plotted as a function of neutron kinetic energy.  The background remaining 
after the subtraction is both flat as a function of energy and consistent with 
zero.  The vertical solid line at 6.0 MeV is the neutron-peak location obtained
from a Gaussian fit, while the vertical dashed line at 6.1 MeV is the expected 
location of the neutron peak based upon the tagger-determined photon energy and
two-body $^{4}$He$(\gamma,n)$ kinematics.  The 100 keV difference is less than 
the 300 keV photon-energy resolution which arises from the physical width of a 
single focal-plane counter.

The true number of neutrons was obtained after a ``stolen-coincidence" 
correction \cite{hornidge99} was applied to the neutron yield.  Stolen 
coincidences occurred when an uncorrelated (random) recoil electron stopped the
focal-plane TDCs prior to a (real) recoil electron correlated in time with a 
neutron event.  A correction was applied to account for these events, which 
would otherwise be missed.  The correction to the neutron yield was 
approximately 5\% for the counting rates employed in the present experiment.

An identical analysis was performed on the empty-target data and demonstrated 
that there was no measurable contribution to the full-target spectra.

\subsection{\label{subsection:cross_section}Cross section}

The laboratory differential cross section for each photon-energy bin was
extracted using

\begin{equation}
\label{equation:dsdw_lab}
\frac{d\sigma}{d\Omega}(E_{\gamma},\theta) =
\frac{Y(E_{\gamma},\theta)}{N_{\gamma}(E_{\gamma}) \cdot \tau_{\rm He} \cdot \Delta \Omega(\theta)}
\end{equation}

\noindent
where $Y(E_{\gamma},\theta)$ was the true neutron yield corrected for
electronic-livetime efficiency, neutron-detection efficiency, neutron-yield
attenuation, and neutron inscattering; $N_{\gamma}(E_{\gamma})$ was the total
number of photons corresponding to a given photon-energy bin corrected for
focal-plane dead-time effects; $\tau_{\rm He}$ was the target thickness 
simulated using a \textsc{geant3} model of the target cell and corrected for 
boiling effects and photon-flux attenuation; and $\Delta \Omega$ was the 
detector geometrical acceptance corrected for extended-target and extended-beam
effects.  The details of the corrections are discussed below.  The resulting 
data are shown in Figure \ref{figure:all_differential_cross_section} and 
presented in Table \ref{table:dsdw_all}.  Figure 
\ref{figure:angle_integrated_cross_section} shows the angle-integrated 
cross-section data, which are presented in Table \ref{table:total}.

\begin{figure}
\begin{center}
\resizebox{0.49\textwidth}{!}{\includegraphics{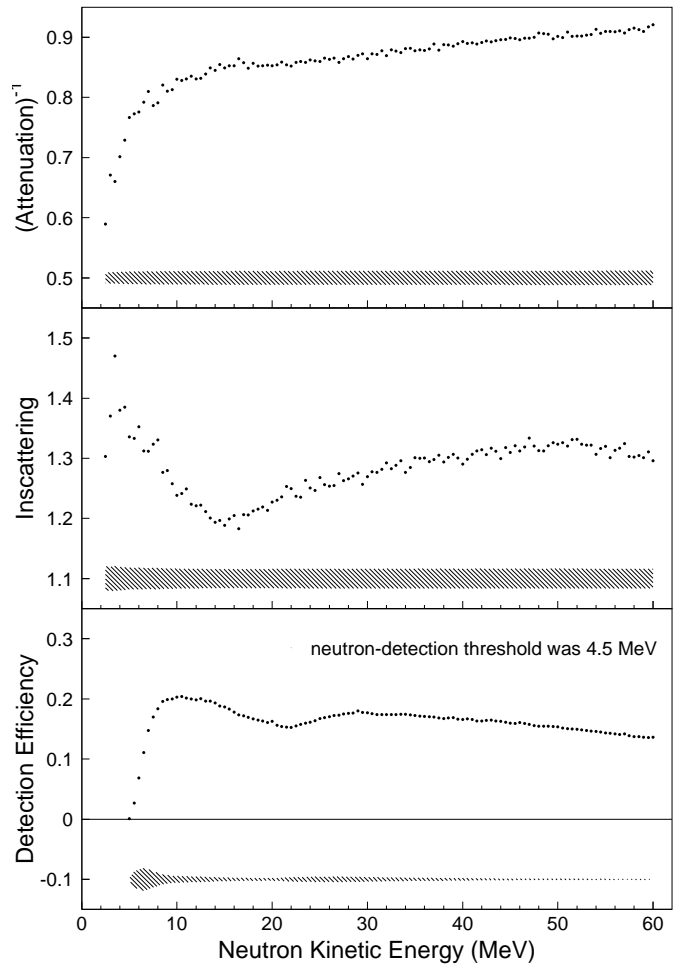}}
\end{center}
\caption{\label{figure:corrections}
The neutron-attenuation, inscattering, and detection-efficiency corrections as 
a function of neutron kinetic energy.  The estimated systematic uncertainties 
in these quantities are represented by the band at the base of each panel.  See
text for further details.}
\end{figure}

\subsubsection{\label{subsubsection:neutron_yield_attenuation}Neutron-yield attenuation}

\begin{figure}
\begin{center}
\resizebox{0.49\textwidth}{!}{\includegraphics{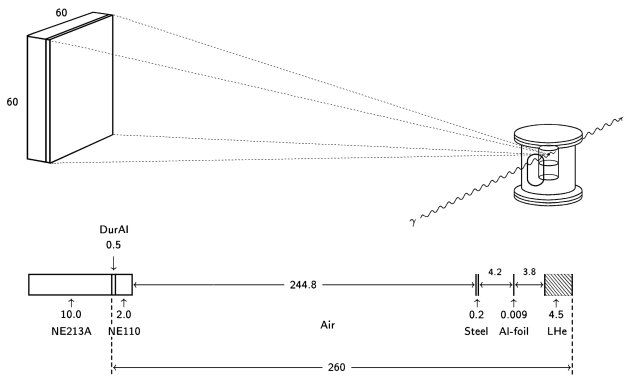}}
\end{center}
\caption{\label{figure:atten_geometry}
The relationship between the photon beam, the target cell, and the detector
aperture is shown in the top panel.  A cross-sectional view of the layers of 
matter between the target and the detector is shown in the bottom panel.  All
dimensions quoted are in cm.  See text for further details.}
\end{figure}

A neutron knocked out of the $^{4}$He nucleus into the acceptance of the 
neutron detectors had to first penetrate significant thicknesses of 
non-detector material before reaching the detector (see 
Figure \ref{figure:atten_geometry}).  Thus, a correction for neutron-yield 
attenuation was necessary.  Neutron absorption in the liquid $^{4}$He, the 
target cell, the target vacuum chamber, the air in the hall, the veto 
scintillator, and the liquid-scintillator canister was determined using a
\textsc{geant3}-based \cite{geant} model of the experiment setup.  The
correction was clearly largest at $T_n$ $<$ 10 MeV, and was dominated by the 
contribution of the veto detector, which was as large as $\sim$10\%.

\subsubsection{\label{subsubsection:neutron_inscattering}Neutron inscattering}

The \textsc{geant3} model was also used to determine the neutron-inscattering 
correction to the data \cite{nilsson06a}.  This effect arose from neutrons
scattering in the materials between and around the target/detector system.
The mass of material surrounding the detector was much greater than the mass
of material which lay in the direct path between the target and the detector.
Thus, the number of neutrons scattering into the detectors (which otherwise 
would have missed) exceeded the number of neutrons scattering out of the 
detectors (which otherwise would have hit).  The Monte-Carlo model of the 
target and detectors was extended to include the neutron-detector lightguides, 
the shielding, and the support tables (recall Figure \ref{figure:ndetector}).  
The model was then embedded in a detailed mockup of the experiment hall, which 
included the concrete floor and ceiling (recall Figure \ref{figure:exphall}).  
Appropriate TOF cuts which took into consideration the measured TOF resolution 
were employed to ensure that only neutrons of the correct energy were 
considered.  The inscattering correction was a strong function of these cuts.  
As shown in the middle panel of Figure \ref{figure:corrections}, this 
correction was also dependent upon neutron energy.

\subsubsection{\label{subsubsection:neutron_detection_efficiency}Neutron-detection efficiency}

The neutron-detection efficiency was determined using the 1979 version of the
\textsc{stanton} code \cite{cecil79}.  This code is based upon neutron 
cross-section data.  The PH thresholds of the neutron-detector cells were set 
as low as possible in hardware to maximize the neutron-detection efficiency.  
However, in the offline analysis, the applied software threshold was varied 
above the hardware value in order to achieve the optimum compromise between 
neutron TOF signal-to-noise ratio and neutron-detection efficiency.  The 
offline software threshold was used because the lower hardware voltage 
threshold had an associated uncertainty due to variations in the shape of the
detected pulse.  Ultimately, in all but the lowest $E_{\gamma}$ bins, the 
average PH threshold employed in the analysis was $\sim$2.0 MeV$_{ee}$, 
corresponding to a neutron energy of $\sim$4.5 MeV, and an average 
neutron-detection efficiency of $\sim$18\% (see Figure 
\ref{figure:corrections}).

Checks of the low-energy predictions made by \textsc{stanton} and 
\textsc{geant3} were made by measuring the neutron-detection efficiency using a
$^{252}$Cf fission-fragment source \cite{karlsson97}.  A detailed description
of the development and testing of \textsc{geant3/stanton}-based simulations of 
neutron detection is given in Ref. \cite{reiter06}.  This testing included the 
measurement of the well-known two-body deuteron photodisintegration cross 
section, which was also performed in previous MAX-lab $(\gamma,n)$ measurements
\cite{annand93,andersson95}.  Unfortunately, time constraints prevented the 
installation of a deuterium target during the present experiment, but all 
MAX-lab $(\gamma,n)$ deuteron-photodisintegration tests of neutron-yield 
corrections have produced $(\gamma,n)$ cross sections which are entirely 
consistent with the generally accepted $(\gamma,p)$ values.

\subsubsection{\label{subsubsection:number_of_photons}Number of photons}

The procedure used for obtaining the number of photons incident upon the target
is presented in detail in Ref. \cite{adler97}.  The incident photon flux for 
each photon-energy bin was determined by counting the number of recoil 
electrons in the tagger focal plane and multiplying the result by the measured 
tagging efficiency, which averaged $\sim$25\%.  Attenuation of the photon flux 
due to atomic processes within the target materials and the liquid helium 
itself \cite{storm70} was also carefully investigated and found to be 
negligible.  The 64 TOF spectra were summed in eight groups of eight tagger 
counters resulting in $\sim$2.5 MeV wide photon-energy bins, each accumulating 
$\sim$10$^{12}$ photons over the course of the measurement.

\subsubsection{\label{subsubsection:geometrical_acceptance}Geometrical acceptance}

The geometrical acceptance of the neutron-detector cell was also determined 
using the \textsc{geant3} simulation.  In this manner, both the extended 
photon-beam profile and the resulting extended-target volume were considered. 
These extended-source effects resulted in a correction of approximately 1\% to 
the $\sim$6 msr point-source acceptance of each cell.

\subsubsection{\label{subsubsection:contamination}Contamination of the two-body peak}
The reaction thresholds for the $\gamma$ $+$ $^{4}$He $\rightarrow$ $^{3}$He + 
$n$ (2bbu) and $\gamma$ $+$ $^{4}$He $\rightarrow$ $^{2}$H $+$ $p$ $+$ $n$ 
(3bbu) are 20.57 and 26.06 MeV, respectively.  Thus, if the neutron energy 
resolution is not sufficient, there may be some contamination of the 2bbu 
signal by 3bbu neutrons.  This is particularly true for the higher 
photon-energy bins which correspond to higher neutron energies, since for a 
fixed target-to-detector flight path, the neutron-energy resolution degrades as 
the velocity increases.  

An upper limit on the level of contamination of our results was determined 
using simulations \cite{annand06} performed within the \textsc{root} 
\cite{root} framework.  These simulations were used to generate 2bbu and 3bbu 
neutrons according to the available kinematic phase space, and considered the 
extended beam, the target, and the detectors in the geometry of the experiment.
With additional smearing to account for the electronic time-pickoff 
uncertainty, the measured widths of the observed 2bbu TOF peaks were reproduced
well.  Equal 2bbu and 3bbu total cross sections were assumed, and the number of
3bbu neutrons which impinged upon the 2bbu neutron-energy integration region 
was determined.  The results are presented in 
Table \ref{table:3bbu_contamination}.

\begin{table}
\caption{\label{table:3bbu_contamination}
A summary of the contamination of the 2bbu yield by neutrons resulting from the
3bbu of $^{4}$He.  The results represent an upper limit determined using equal 
2bbu and 3bbu total cross sections.  See text for further details.}
\begin{ruledtabular}
\begin{tabular}{rc}
$E_{\gamma}$ &    $<$contamination$>$ \\
       (MeV) &                   (\%) \\
\hline
        38.8 & \hspace*{-2.5mm}$<$0.1 \\
        40.6 & \hspace*{-2.5mm}$<$0.1 \\
        51.5 &                    1.0 \\
        53.6 &                    1.5 \\
        56.0 &                    2.0 \\
        58.4 &                    2.6 \\
        63.6 &                    4.1 \\
        68.5 &                    5.0 \\
\end{tabular}
\end{ruledtabular}
\end{table}

Cross-section data for the 3bbu reaction in the present energy range are 
sparce,  but the available evidence \cite{quaglioni04} suggests that the 3bbu 
cross section is less than that for 2bbu.  Thus, we believe that the results 
presented in Table \ref{table:3bbu_contamination} represent an upper limit on 
the contamination.  We do not correct our data for this effect.

\subsubsection{\label{subsubsection:systematic_uncertainties}Systematic uncertainties}

\begin{figure}
\begin{center}
\resizebox{0.47\textwidth}{!}{\includegraphics{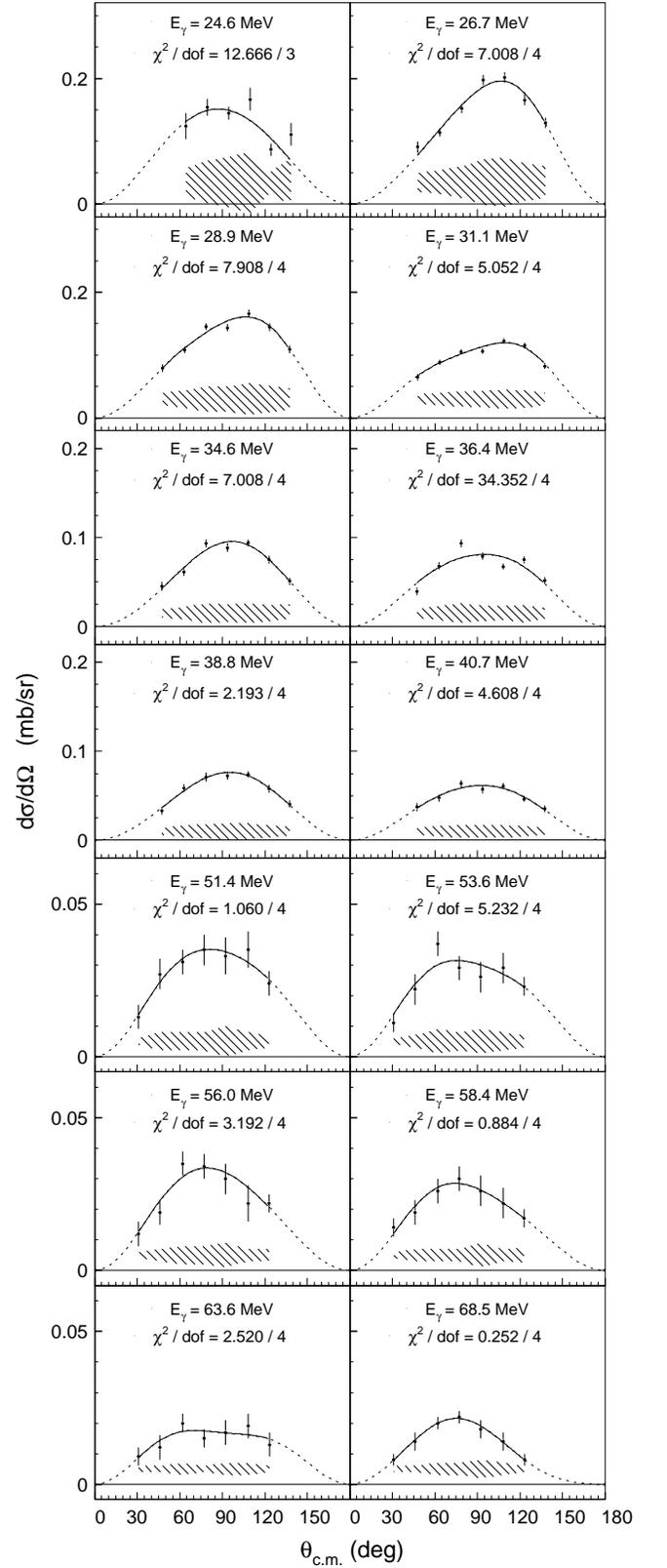}}
\end{center}
\caption{\label{figure:all_differential_cross_section}
c.m. angular distributions for the $^{4}$He$(\gamma,n)$ reaction from 23 $<$ 
E$_{\gamma}$ $<$ 70 MeV.  Error bars are the statistical uncertainties, while 
the systematic uncertainties are represented by the bands at the base of each 
panel.  The solid lines are the fitted functions, while the dashed lines are 
the fitted functions extrapolated to zero at $\theta_{\rm c.m.}$ $=$ (0,180) 
$\deg$.  See text for further details and Table \ref{table:dsdw_all} for 
numerical values.}
\end{figure}

The systematic uncertainty in the measurement was dominated by the systematic 
uncertainty in the neutron-detection efficiency, which ranged from $\sim$5\% 
for $E_{\gamma}$ $=$ 68.5 MeV to $\sim$26\% for $E_{\gamma}$ $=$ 24.6 MeV.  
Other large sources of uncertainty were the neutron-inscattering correction 
($\sim$9\%), the neutron-yield attenuation correction ($\sim$6\%), and the 
number of photons (a combination of the tagger focal-plane livetime and the
tagging efficiency; $\sim$4\%).  A summary of the systematic uncertainties is 
presented in Table \ref{table:systematic_uncertainties}.  The systematic 
uncertainties associated with each of the individual cross-section data points 
are presented in Tables \ref{table:dsdw_all} and \ref{table:total}.  See also 
the discussion of the additional systematic uncertainties arising from the 
analysis of the angular distributions presented in 
Section \ref{subsection:angular_distributions}, and the uncertainty bands shown
in Figures \ref{figure:all_differential_cross_section} and 
\ref{figure:angle_integrated_cross_section}.

\begin{table}
\caption{\label{table:systematic_uncertainties}
A summary of the correction factors applied to the cross-section data together
with systematic uncertainties.  In the case of the kinematic-dependent
corrections, average values for the correction and the uncertainty are stated.}
\begin{ruledtabular}
\begin{tabular}{rcc}
kinematic-dependent quantity & $<$value$>$ &      $<$uncertainty$>$ \\
\hline
neutron-detection efficiency &        0.20 &                    8\% \\
        neutron-inscattering &        1.25 &                    9\% \\
   neutron-yield attenuation &        0.85 &                    6\% \\
 tagger focal-plane livetime &        0.95 &                    2\% \\
   neutron-detector livetime &        0.50 &                    1\% \\
     photon-beam attenuation &  (see text) & \hspace*{-2.5mm}$<$1\% \\
\hline
              scale quantity &       value &            uncertainty \\
\hline
          tagging efficiency &        0.25 &                    3\% \\
      geometrical acceptance &  (see text) &                    2\% \\
              target density &  (see text) &                    2\% \\
  particle misidentification &  (see text) & \hspace*{-2.5mm}$<$1\% \\
\end{tabular}
\end{ruledtabular}
\end{table}

\section{\label{section:results}Results}

\subsection{\label{subsection:calculations}The calculations}

In this Section, we compare our data to Recoil-Corrected Continuum Shell Model 
(RCCSM) calculations \cite{halderson81,halderson04}, a Resonating Group Method 
(RGM) calculation \cite{wachter88}, and an Effective Interaction Hyperspherical
Harmonic (EIHH) Expansion calculation \cite{quaglioni04}.  Note that the 
authors of the RCCSM and RGM calculations originally presented their results in
the form of Legendre coefficients for the $^{3}$He$(n,\gamma)$ reaction 
expressed as a function of Center-of-Mass (c.m.) proton energy corresponding to
the $^{3}$H$(p,\gamma)$ reaction, while the authors of the EIHH calculations 
present their results as a function of photon energy.  These calculations 
consider only the two-fragment photodisintegration of $^{4}$He into the 
($n+^{3}$He) final state.

The RCCSM calculations were performed using a continuum shell-model framework
in the ($1p1h$) approximation where the transition-matrix elements of the M1
and the (spin-independent) M2 multipole operators vanished.  Target-recoil
corrections were applied.  The effective nucleon-nucleon (NN) interaction 
included central, spin-orbit, and tensor components in addition to the Coulomb 
force.  Perturbation theory was used to compute matrix elements for the 
multipoles.  The multipole operators were calculated in the long-wavelength 
limit.  After corrections applied for spurious c.m. excitations, these 
calculations were essentially equivalent to the multichannel microscopic RGM 
calculations described below.  Note that the newer RCCSM 
calculation \cite{halderson04} expanded the model space of the earlier 
calculation \cite{halderson81} to include more reaction channels and all 
$p$-shell nuclei.

The multichannel microscopic RGM calculations were performed using a
semi-realistic NN force similar to the one detailed above.  The variational 
principle was used to determine the scattering wave functions.  Radiative 
processes were treated within the Born Approximation, and the electromagnetic 
transition operators were again taken in the long-wavelength limit.  Angular 
momenta up to $L=2$ were allowed in the relative motion of the fragments.  To 
the knowledge of the authors, further development of the RGM framework for 
photonuclear processes has ceased.

The EIHH calculation used a correlated hyperspherical expansion of basis 
states, with final-state interactions accounted for in a rigorous manner using 
the Lorentz Integral Transform Method (which circumvents the calculation of 
continuum wave functions).

\vspace*{4mm}

\subsection{\label{subsection:angular_distributions}Angular distributions}

The angular distributions measured at each photon energy were converted from
the laboratory to the c.m. frame.  In the two angular analyses we present 
below (and similar to analyses of complementary $^{4}$He$(\gamma,p)$ 
angular-distribution data \cite{jones91,calarco03}), we constrained our angular
distributions to vanish at $\theta_{\rm c.m.}$ $=$ (0,180) $\deg$.  We note 
that Weller {\it{et al.}~}\cite{weller82} claim non-zero interfering E1 $S=1$ 
strength, which results in a non-vanishing angular distribution at 
$\theta_{\rm c.m.}$ $=$ (0,180) $\deg$.  However, the present data do not have 
the precision and angular range necessary to investigate this small effect.

The systematic uncertainties in the angular distribution coefficients were
estimated from the systematic uncertainties in the angular distributions (see 
Section \ref{subsubsection:systematic_uncertainties}).  Three extreme scenarios
were considered, where the systematic uncertainty in the differential 
cross-section data would have a maximal effect upon the extracted 
angular-distribution coefficient.  The first scenario involved shifting all the
differential cross-section data points in an angular distribution either up or 
down in unison by their associated systematic uncertainty.  The second scenario
involved shifts of the same magnitude, but not in unison.  Rather, they were
made to either emphasize or de-emphasize the degree of forward/backward 
asymmetry in the angular distribution.  The third scenario again involved 
shifts of the same magnitude, but this time to either emphasize or de-emphasize
the peaking of the angular distribution at $\theta_{\rm c.m.}$ $\sim$ 
90$^{\circ}$.  These three extreme sets of angular distributions were fitted
as described in the sections below.  The resulting systematic uncertainties
were taken as the average spread in the value of the derived coefficients, and 
are displayed as error bands in Figures \ref{figure:abc_coeffs} and
\ref{figure:legendre_coefficients}.

\subsubsection{\label{subsubsection:abc}The Transition-coefficient Approach}

In the context of the Transition-coefficient Approach, the c.m. angular 
distributions were fitted using
\begin{widetext}
\begin{equation}
\label{equation:dsdw_cm_abc}
\frac{d\sigma}{d\Omega}(\theta_{\rm c.m.}) = 
\alpha~\left\{\sin^{2}(\theta_{\rm c.m.})\left[1 + \beta \cos(\theta_{\rm c.m.}) + \gamma \cos^{2}(\theta_{\rm c.m.})\right] + \delta + \epsilon \cos(\theta_{\rm c.m.})~\right\}.
\end{equation}
\end{widetext}
This expansion assumes that the photon multipolarities are restricted to E1,
E2, and M1, and that the nuclear matrix elements of the E-multipoles to final 
states with a channel spin of unity are negligible \cite{jones91}.  Under these
assumptions, $\alpha$ arises from the incoherent sum of the E1, E2, and M1 
multipoles, $\beta$ is due to the interference of the E1 and E2 multipoles, 
$\gamma$ results from the E2 multipole, $\delta$ arises from the M1 multipole, 
and $\epsilon$ is vanishingly small.  As previously mentioned, in this 
analysis, the angular distributions were constrained to vanish at 
$\theta_{\rm c.m.}$ $=$ (0,180) $\deg$ -- in this case, by forcing the 
$\delta$ and $\epsilon$ coefficients to be zero.

\begin{figure}
\begin{center}
\resizebox{0.47\textwidth}{!}{\includegraphics{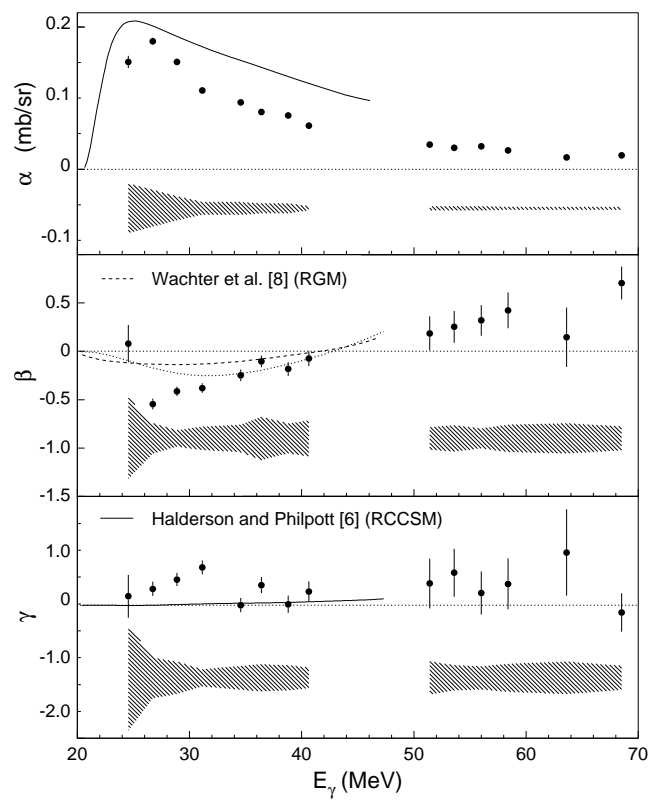}}
\end{center}
\caption{\label{figure:abc_coeffs}
The $\alpha$, $\beta$, and $\gamma$ coefficients:  present data -- filled
circles; earlier RCCSM calculations \cite{halderson81} -- solid lines; RGM 
calculation \cite{wachter88} -- dashed line.  Error bars are the statistical 
uncertainties, while the systematic uncertainties are represented by the bands 
at the base of each panel.  See text for further details.}
\end{figure}

Figure \ref{figure:abc_coeffs} presents the $\alpha$, $\beta$, and $\gamma$
coefficients (filled circles) as a function of photon energy.  The values are 
summarized in Table \ref{table:table_abc}.  Error bars are the statistical 
uncertainties, while the systematic uncertainties are represented by the bands 
at the base of each panel.  Also shown are earlier RCCSM \cite{halderson81} and 
RGM \cite{wachter88} calculations.  Angular distributions were not published
in Refs. \cite{halderson04} (newer RCCSM) or \cite{quaglioni04} (EIHH).

As shown, the data basically follow the trends predicted by the calculations.  
At the lower photon energies where the E1 multipole is completely dominant, 
the $\alpha$-coefficient data have a clear resonant structure peaking at 
$E_{\gamma}$ $\sim$ 28 MeV.  The earlier RCCSM calculation tends to 
overestimate the data, but also shows resonant structure peaking at 
$E_{\gamma}$ $\sim$ 25 MeV.  The energy dependence of the $\beta$-coefficient 
data is reasonably consistent with both the earlier RCCSM and the RGM 
predictions, given the systematic uncertainties for $E_{\gamma}$ $<$ 26 MeV.  
Similarly, there is no significant disagreement between the present 
$\gamma$-coefficient data and the earlier RCCSM calculation when uncertainties 
are considered.  At higher photon energies, E2 strength is expected to become 
more important.  Unfortunately, the calculations do not cover the range of 
the higher-energy data.  That said, these data do appear to be consistent with 
the energy-extrapolated trends of both the lower-energy data and the 
calculations.

\subsubsection{\label{subsubsection:legendre}The Legendre Approach}

In the context of the Legendre Approach, the c.m. angular distributions were 
fitted using

\begin{equation}
\label{equation:dsdw_cm_legendre}
\frac{d\sigma}{d\Omega}(\theta_{\rm c.m.}) =
A_{0}\left[1 + \sum_{n=1}^{4} a_{n}P_{n}(\cos(\theta_{\rm c.m.}))\right].
\end{equation}
The angular distributions were constrained to vanish at $\theta_{\rm c.m.}$ 
$=$ (0,180) $\deg$ by enforcing the constraints $a_{1} = -a_{3}$ and 
$1 + a_{2} + a_{4} = 0$ (equivalent to the $\delta = \epsilon = 0$
constraints used in the Transition-coefficient Approach).

\begin{figure}
\begin{center}
\resizebox{0.47\textwidth}{!}{\includegraphics{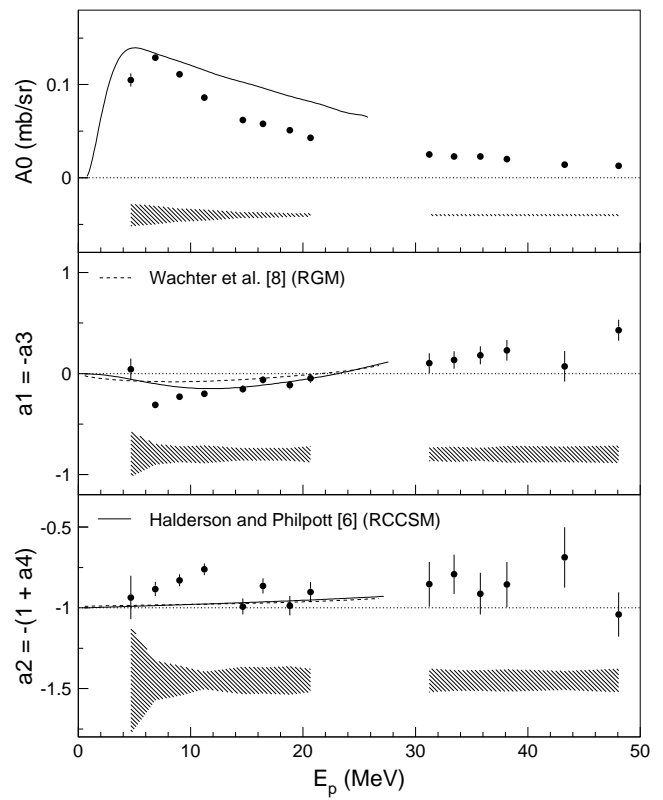}}
\end{center}
\caption{\label{figure:legendre_coefficients}
The Legendre coefficients:  present data -- filled circles; earlier RCCSM 
calculations \cite{halderson81} -- solid lines; RGM calculation
\cite{wachter88} -- dashed line.  Error bars are the statistical 
uncertainties, while the systematic uncertainties are represented by the 
bands at the base of each panel.  See text for further details.}
\end{figure}

Figure \ref{figure:legendre_coefficients} presents the $A_0$ and $a_1$--$a_4$
coefficients (filled circles).  Values are summarized in 
Table \ref{table:legendre_coefficients}.  Error bars are the statistical 
uncertainties, while the systematic uncertainties are represented by the bands 
at the base of each panel.  Also shown are earlier RCCSM \cite{halderson81} and
RGM \cite{wachter88} calculations for the $^{3}$He$(n,\gamma)$ reaction.  In
keeping with the convention chosen by the authors of these theoretical works, 
our data and their calculations have been plotted as a function of c.m. 
proton energy for the $^{3}$H$(p,\gamma)$ reaction.

As shown again, the data largely reproduce the trends predicted by the 
calculations.  At lower energies, the E1 multipole is completely
dominant and the $A_0$ coefficient has a clear resonant structure peaking at 
$E_{p}$ $\sim$ 7 MeV.  The earlier RCCSM calculation tends to overestimate 
these data, but also has the resonant structure peaking at $E_{p}$ $\sim$ 
6 MeV.  The energy dependence of the $a_1 = -a_3$ data is reasonably 
consistent with both the earlier RCCSM and RGM predictions, given the 
systematic uncertainties for $E_p$ $<$ 8 MeV.  Similarly, there is no 
significant disagreement between the present $a_2 = -(1 + a_4)$ data and the 
calculations.  Finally, while the calculations again do not cover the range of 
the higher-energy data where the E2 strength is expected to become more 
important, these data do appear to be consistent with the energy-dependent 
trends of both the lower-energy data and the calculations.

\subsection{\label{subsection:angle_integrated}Angle-integrated cross section}

\begin{figure}
\begin{center}
\resizebox{0.44\textwidth}{!}{\includegraphics{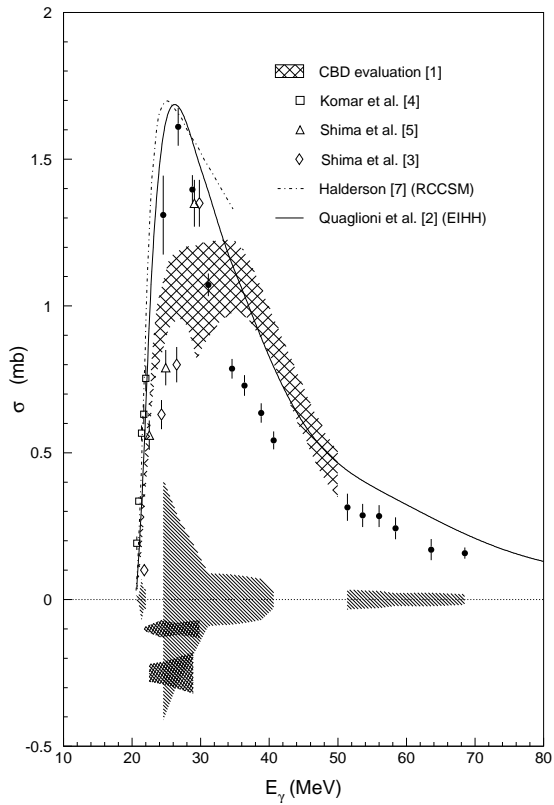}}
\end{center}
\caption{\label{figure:angle_integrated_cross_section}
The angle-integrated $^{4}$He$(\gamma,n)$ cross section:  present data --
filled circles; CBD evaluation \cite{calarco83} -- hatched band; RCCSM
calculation \cite{halderson04} -- dashed-dotted line; and EIHH calculation
\cite{quaglioni04} -- solid line.  Error bars are the statistical
uncertainties, while the systematic uncertainties associated with each of the 
data sets are represented by the bands at the base of the panel.  See text for
further details.}
\end{figure}

Figure \ref{figure:angle_integrated_cross_section} presents the 
angle-integrated cross-section data (filled circles).  Also shown are the CBD 
evaluation \cite{calarco83}, data from a $^{3}$He$(n,\gamma)$ 
measurement \cite{komar93}, data from $^{4}$He$(\gamma,^{3}$He$)$ active-target
measurements \cite{shima01,shima05}, the newer RCCSM 
calculation \cite{halderson04}, and the EIHH calculation \cite{quaglioni04}.  
Note that both calculations employ the semi-realistic MTI-III 
potential \cite{malfliet69}.  Error bars show the statistical uncertainties, 
while the systematic uncertainties are represented by the bands at the base of 
the panel.  For clarity, the systematic uncertainties in the data from 
Refs. \cite{shima05,shima01} have been centered at $-$0.1 and $-$0.25, 
respectively.  Also for clarity, the small uncertainty in the EIHH calculation 
for the photon-energy region between 2bbu threshold at $E_{\gamma}$ $=$ 20.6 
MeV and 3bbu threshold at $E_{\gamma}$ $=$ 26.1 MeV discussed in 
Ref. \cite{quaglioni04} is not shown here.

The present $^{4}$He$(\gamma,n)$ angle-integrated cross-section data has a 
clear resonant structure which peaks at $E_{\gamma}$ $\sim$ 28 MeV.  On 
average, these data are approximately 7\% larger than those which result from 
simply scaling our projected $\theta_{\rm c.m.}$ $=$ 90 $\deg$ results by 
$8\pi/3$.  Although data are lacking for 42 $<$ $E_{\gamma}$ $<$ 50 MeV,
there is no apparent discontinuity in this region.  Furthermore, the present 
data extrapolate smoothly to the lower-energy data of Ref. \cite{komar93}.  
Conversely, the data of Refs. \cite{shima01,shima05} below 25 MeV are at odds 
with all other data, the calculations, and the CBD evaluation, although it is 
in good agreement with the present experiment near 30 MeV.  Both the RCCSM and 
EIHH calculations are in good agreement with the present data and those of 
Ref. \cite{komar93} up to the resonant peak at $E_{\gamma}$ $\sim$ 28 MeV.  At 
higher energies, both calculations tend to overpredict the data.  Nevertheless,
the EIHH calculation follows the general shape of the excitation function up to
$E_{\gamma}$ $\sim$ 70 MeV reasonably well.  Development of the EIHH formalism 
continues \cite{gazit05,barnea06} so that the total photoabsorption may now be 
calculated using the Argonne V18 NN potential in conjunction with the Urbana IX
3N potential, and we anticipate new predictions for the partial two-body 
photodisintegration channels in the near future.

\section{\label{section:summary}Summary and conclusions}

In summary, $\frac{d\sigma}{d\Omega}(E_{\gamma},\theta)$ for the 
$^{4}$He$(\gamma,n)$ reaction have been measured with tagged photons and 
compared to other available measurements and calculations.  The energy 
dependence of the transition coefficients $\alpha$, $\beta$, and $\gamma$ as 
well as the Legendre coefficients $A_0$, $a_1$, and $a_2$ extracted from the 
angular distributions agrees reasonably well with trends predicted by earlier 
RCCSM \cite{halderson81} and RGM \cite{wachter88} calculations.  The marked 
resonant behavior of the present angle-integrated cross section, peaking at 
$E_{\gamma}$ $\sim$ 28 MeV, is in good agreement with newer 
RCCSM \cite{halderson04} and EIHH \cite{quaglioni04} calculations as well as 
capture data \cite{komar93} which extend close to the $(\gamma,n)$ threshold.  
This behavior disagrees with an evaluation of $(\gamma,n)$ 
data \cite{calarco83} made in 1983, and recent active-target 
data \cite{shima01,shima05}.

\begin{acknowledgments}

The authors acknowledge the outstanding support of the MAX-lab staff which made
this experiment successful.  We also wish to thank Sofia Quaglioni, Winfried 
Leidemann, and Giuseppina Orlandini (University of Trento, Italy), John Calarco
(University of New Hampshire, USA), Victor Efros (Kurchatov Institute, Russia),
Gerald Feldman (The George Washington University, USA), Dean Halderson (Western
Michigan University, USA), Andreas Reiter (University of Glasgow, Scotland), 
and Brad Sawatzky (University of Virginia, USA) for valuable discussions. B.N. 
wishes to thank Margareta S\"{o}derholm and Ralph Hagberg for their unwavering 
support.  The Lund group acknowledges the financial support of the Swedish 
Research Council, the Knut and Alice Wallenberg Foundation, the Crafoord 
Foundation, the Swedish Institute, the Wenner-Gren Foundation, and the Royal 
Swedish Academy of Sciences.  The Glasgow group acknowledges the financial 
support of the UK Engineering and Physical Sciences Research Council.

\end{acknowledgments}


\appendix

\section{\label{section:cross_section_data}Data tables}

\subsection{\label{subsection:differential_cross_section_data}Differential cross-section data}

A summary of the differential cross-section data is presented in Table
\ref{table:dsdw_all}.

\begin{table}
\caption{\label{table:dsdw_all}
The differential cross-section data (mb/sr) for the $^{4}$He$(\gamma,n)$ 
reaction shown in Fig. \ref{figure:all_differential_cross_section}.  The first 
uncertainty is statistical and the second uncertainty is systematic.}
\begin{ruledtabular}
\begin{tabular}{rc|rc}
$\theta_{\rm c.m.}$ &   d$\sigma$/d$\Omega_{\rm c.m.}$ & $\theta_{\rm c.m.}$ &   d$\sigma$/d$\Omega_{\rm c.m.}$ \\
         ($\deg$) &                        (mb/sr) &          ($\deg$) &                        (mb/sr) \\
\hline
    \multicolumn{2}{c|}{$E_{\gamma}$ $=$ 24.6 MeV} &      \multicolumn{2}{c}{$E_{\gamma}$ $=$ 26.7 MeV} \\
\hline
                  &                                &              47.9 &  0.091 $\pm$ 0.009 $\pm$ 0.015 \\
             64.1 &  0.124 $\pm$ 0.020 $\pm$ 0.023 &              63.6 &  0.114 $\pm$ 0.007 $\pm$ 0.019 \\
             79.6 &  0.154 $\pm$ 0.014 $\pm$ 0.034 &              79.0 &  0.152 $\pm$ 0.007 $\pm$ 0.025 \\
             94.8 &  0.145 $\pm$ 0.010 $\pm$ 0.040 &              94.2 &  0.197 $\pm$ 0.008 $\pm$ 0.036 \\
            109.6 &  0.167 $\pm$ 0.018 $\pm$ 0.049 &             109.0 &  0.202 $\pm$ 0.008 $\pm$ 0.039 \\
            124.1 &  0.087 $\pm$ 0.010 $\pm$ 0.017 &             123.6 &  0.166 $\pm$ 0.009 $\pm$ 0.030 \\
            138.4 &  0.111 $\pm$ 0.018 $\pm$ 0.038 &             137.9 &  0.130 $\pm$ 0.008 $\pm$ 0.025 \\
\hline
    \multicolumn{2}{c|}{$E_{\gamma}$ $=$ 28.9 MeV} &      \multicolumn{2}{c}{$E_{\gamma}$ $=$ 31.1 MeV} \\
\hline
             47.7 &  0.079 $\pm$ 0.006 $\pm$ 0.010 &              47.6 &  0.065 $\pm$ 0.005 $\pm$ 0.008 \\
             63.4 &  0.108 $\pm$ 0.005 $\pm$ 0.014 &              63.2 &  0.089 $\pm$ 0.004 $\pm$ 0.010 \\
             78.7 &  0.145 $\pm$ 0.005 $\pm$ 0.020 &              78.6 &  0.105 $\pm$ 0.004 $\pm$ 0.012 \\
             93.9 &  0.144 $\pm$ 0.006 $\pm$ 0.020 &              93.7 &  0.106 $\pm$ 0.004 $\pm$ 0.013 \\
            108.7 &  0.166 $\pm$ 0.006 $\pm$ 0.025 &             108.6 &  0.123 $\pm$ 0.004 $\pm$ 0.015 \\
            123.4 &  0.144 $\pm$ 0.006 $\pm$ 0.021 &             123.2 &  0.115 $\pm$ 0.005 $\pm$ 0.014 \\
            137.7 &  0.109 $\pm$ 0.006 $\pm$ 0.017 &             137.6 &  0.082 $\pm$ 0.004 $\pm$ 0.010 \\
\hline 
    \multicolumn{2}{c|}{$E_{\gamma}$ $=$ 34.6 MeV} &      \multicolumn{2}{c}{$E_{\gamma}$ $=$ 36.4 MeV} \\
\hline
             47.5 &  0.045 $\pm$ 0.005 $\pm$ 0.005 &              47.5 &  0.040 $\pm$ 0.004 $\pm$ 0.005 \\
             63.1 &  0.061 $\pm$ 0.004 $\pm$ 0.007 &              63.0 &  0.068 $\pm$ 0.004 $\pm$ 0.008 \\
             78.5 &  0.093 $\pm$ 0.004 $\pm$ 0.011 &              78.4 &  0.093 $\pm$ 0.005 $\pm$ 0.011 \\
             93.6 &  0.089 $\pm$ 0.004 $\pm$ 0.010 &              93.5 &  0.079 $\pm$ 0.004 $\pm$ 0.009 \\
            108.5 &  0.094 $\pm$ 0.004 $\pm$ 0.011 &             108.5 &  0.067 $\pm$ 0.003 $\pm$ 0.008 \\
            123.1 &  0.075 $\pm$ 0.004 $\pm$ 0.009 &             123.1 &  0.075 $\pm$ 0.004 $\pm$ 0.009 \\
            137.5 &  0.051 $\pm$ 0.004 $\pm$ 0.006 &             137.5 &  0.052 $\pm$ 0.004 $\pm$ 0.006 \\
\hline
    \multicolumn{2}{c|}{$E_{\gamma}$ $=$ 38.8 MeV} &      \multicolumn{2}{c}{$E_{\gamma}$ $=$ 40.7 MeV} \\
\hline
             47.5 &  0.033 $\pm$ 0.004 $\pm$ 0.004 &              47.5 &  0.037 $\pm$ 0.004 $\pm$ 0.004 \\
             63.1 &  0.059 $\pm$ 0.004 $\pm$ 0.007 &              63.0 &  0.047 $\pm$ 0.004 $\pm$ 0.006 \\
             78.4 &  0.071 $\pm$ 0.005 $\pm$ 0.008 &              78.4 &  0.064 $\pm$ 0.004 $\pm$ 0.007 \\
             93.5 &  0.073 $\pm$ 0.004 $\pm$ 0.009 &              93.5 &  0.057 $\pm$ 0.004 $\pm$ 0.007 \\
            108.4 &  0.074 $\pm$ 0.004 $\pm$ 0.009 &             108.4 &  0.061 $\pm$ 0.003 $\pm$ 0.007 \\
            123.0 &  0.058 $\pm$ 0.004 $\pm$ 0.007 &             123.0 &  0.047 $\pm$ 0.003 $\pm$ 0.006 \\
            137.5 &  0.041 $\pm$ 0.004 $\pm$ 0.005 &             137.5 &  0.035 $\pm$ 0.003 $\pm$ 0.004 \\
\hline
    \multicolumn{2}{c|}{$E_{\gamma}$ $=$ 51.4 MeV} &      \multicolumn{2}{c}{$E_{\gamma}$ $=$ 53.6 MeV} \\
\hline
             31.0 &  0.013 $\pm$ 0.004 $\pm$ 0.001 &              31.0 &  0.011 $\pm$ 0.003 $\pm$ 0.001 \\
             45.9 &  0.027 $\pm$ 0.005 $\pm$ 0.003 &              45.9 &  0.022 $\pm$ 0.005 $\pm$ 0.002 \\
             62.1 &  0.031 $\pm$ 0.004 $\pm$ 0.003 &              62.1 &  0.037 $\pm$ 0.004 $\pm$ 0.004 \\
             77.0 &  0.035 $\pm$ 0.005 $\pm$ 0.003 &              77.0 &  0.029 $\pm$ 0.004 $\pm$ 0.003 \\
             92.1 &  0.033 $\pm$ 0.006 $\pm$ 0.005 &              92.1 &  0.026 $\pm$ 0.005 $\pm$ 0.004 \\
            108.3 &  0.035 $\pm$ 0.006 $\pm$ 0.003 &             108.3 &  0.029 $\pm$ 0.005 $\pm$ 0.003 \\
            123.1 &  0.024 $\pm$ 0.004 $\pm$ 0.002 &             123.1 &  0.023 $\pm$ 0.003 $\pm$ 0.002 \\
\hline
    \multicolumn{2}{c|}{$E_{\gamma}$ $=$ 56.0 MeV} &      \multicolumn{2}{c}{$E_{\gamma}$ $=$ 58.4 MeV} \\
\hline
             31.0 &  0.012 $\pm$ 0.004 $\pm$ 0.001 &              31.0 &  0.014 $\pm$ 0.003 $\pm$ 0.001 \\
             45.9 &  0.019 $\pm$ 0.004 $\pm$ 0.002 &              45.9 &  0.019 $\pm$ 0.004 $\pm$ 0.002 \\
             62.1 &  0.035 $\pm$ 0.004 $\pm$ 0.003 &              62.1 &  0.026 $\pm$ 0.004 $\pm$ 0.002 \\
             77.0 &  0.034 $\pm$ 0.004 $\pm$ 0.003 &              77.0 &  0.030 $\pm$ 0.004 $\pm$ 0.002 \\
             92.1 &  0.030 $\pm$ 0.005 $\pm$ 0.004 &              92.1 &  0.026 $\pm$ 0.005 $\pm$ 0.004 \\
            108.3 &  0.022 $\pm$ 0.006 $\pm$ 0.002 &             108.3 &  0.022 $\pm$ 0.005 $\pm$ 0.002 \\
            123.1 &  0.022 $\pm$ 0.003 $\pm$ 0.002 &             123.2 &  0.017 $\pm$ 0.003 $\pm$ 0.003 \\
\hline
    \multicolumn{2}{c|}{$E_{\gamma}$ $=$ 63.6 MeV} &      \multicolumn{2}{c}{$E_{\gamma}$ $=$ 68.5 MeV} \\
\hline
             31.0 &  0.009 $\pm$ 0.003 $\pm$ 0.001 &              31.0 &  0.008 $\pm$ 0.002 $\pm$ 0.001 \\
             45.9 &  0.012 $\pm$ 0.004 $\pm$ 0.001 &              45.9 &  0.014 $\pm$ 0.003 $\pm$ 0.001 \\
             62.1 &  0.020 $\pm$ 0.003 $\pm$ 0.002 &              62.1 &  0.020 $\pm$ 0.002 $\pm$ 0.002 \\
             77.0 &  0.015 $\pm$ 0.003 $\pm$ 0.001 &              77.0 &  0.022 $\pm$ 0.002 $\pm$ 0.002 \\
             92.1 &  0.017 $\pm$ 0.004 $\pm$ 0.002 &              92.1 &  0.018 $\pm$ 0.003 $\pm$ 0.003 \\
            108.3 &  0.019 $\pm$ 0.004 $\pm$ 0.002 &             108.3 &  0.014 $\pm$ 0.003 $\pm$ 0.002 \\
            123.2 &  0.013 $\pm$ 0.004 $\pm$ 0.001 &             123.3 &  0.008 $\pm$ 0.002 $\pm$ 0.001 \\
\end{tabular}
\end{ruledtabular}
\end{table}

\subsection{\label{subsection:angular_distribution coefficients}Angular-distribution coefficients}

A summary of the angular-distribution coefficients is presented in
Tables \ref{table:table_abc} and \ref{table:legendre_coefficients}.

\begin{table*}
\caption{\label{table:table_abc}
A summary of the coefficients $\alpha$, $\beta$, and $\gamma$ extracted from 
the data.  The first uncertainty is statistical and the second uncertainty is 
systematic.  See also Figure \ref{figure:abc_coeffs}.}
\begin{ruledtabular}
\begin{footnotesize}
\begin{tabular}{crrr}
$E_{\gamma}$ &                      $\alpha$ &                          $\beta$ &                         $\gamma$ \\
       (MeV) &                       (mb/sr) &                                  &                                  \\
\hline
        24.6 & 0.151 $\pm$ 0.009 $\pm$ 0.035 &    0.081 $\pm$ 0.193 $\pm$ 0.418 &    0.186 $\pm$ 0.440 $\pm$ 1.093 \\
        26.7 & 0.180 $\pm$ 0.005 $\pm$ 0.026 & $-$0.545 $\pm$ 0.055 $\pm$ 0.158 &    0.341 $\pm$ 0.149 $\pm$ 0.421 \\
        28.9 & 0.151 $\pm$ 0.004 $\pm$ 0.017 & $-$0.414 $\pm$ 0.048 $\pm$ 0.085 &    0.533 $\pm$ 0.134 $\pm$ 0.335 \\
        31.1 & 0.111 $\pm$ 0.003 $\pm$ 0.009 & $-$0.380 $\pm$ 0.050 $\pm$ 0.121 &    0.785 $\pm$ 0.143 $\pm$ 0.178 \\
        34.6 & 0.094 $\pm$ 0.003 $\pm$ 0.009 & $-$0.249 $\pm$ 0.061 $\pm$ 0.141 &    0.004 $\pm$ 0.148 $\pm$ 0.246 \\
        36.4 & 0.080 $\pm$ 0.003 $\pm$ 0.007 & $-$0.104 $\pm$ 0.060 $\pm$ 0.224 &    0.419 $\pm$ 0.166 $\pm$ 0.280 \\
        38.8 & 0.076 $\pm$ 0.003 $\pm$ 0.003 & $-$0.183 $\pm$ 0.071 $\pm$ 0.153 &    0.021 $\pm$ 0.179 $\pm$ 0.260 \\
        40.7 & 0.061 $\pm$ 0.003 $\pm$ 0.003 & $-$0.075 $\pm$ 0.077 $\pm$ 0.185 &    0.287 $\pm$ 0.206 $\pm$ 0.215 \\
        51.4 & 0.034 $\pm$ 0.004 $\pm$ 0.003 &    0.186 $\pm$ 0.176 $\pm$ 0.117 &    0.450 $\pm$ 0.516 $\pm$ 0.344 \\
        53.6 & 0.030 $\pm$ 0.003 $\pm$ 0.003 &    0.253 $\pm$ 0.165 $\pm$ 0.136 &    0.672 $\pm$ 0.494 $\pm$ 0.262 \\
        56.0 & 0.032 $\pm$ 0.003 $\pm$ 0.003 &    0.319 $\pm$ 0.157 $\pm$ 0.103 &    0.254 $\pm$ 0.443 $\pm$ 0.239 \\
        58.4 & 0.027 $\pm$ 0.003 $\pm$ 0.002 &    0.423 $\pm$ 0.185 $\pm$ 0.141 &    0.443 $\pm$ 0.526 $\pm$ 0.289 \\
        63.6 & 0.017 $\pm$ 0.003 $\pm$ 0.002 &    0.146 $\pm$ 0.308 $\pm$ 0.156 &    1.091 $\pm$ 0.893 $\pm$ 0.336 \\
        68.5 & 0.019 $\pm$ 0.002 $\pm$ 0.002 &    0.706 $\pm$ 0.170 $\pm$ 0.124 & $-$0.151 $\pm$ 0.398 $\pm$ 0.250 \\
\end{tabular}
\end{footnotesize}
\end{ruledtabular}
\end{table*}

\begin{table*}
\caption{\label{table:legendre_coefficients}
A summary of the Legendre coefficients extracted from the data.  The first 
uncertainty is statistical and the second uncertainty is systematic.  See also 
Figure \ref{figure:legendre_coefficients}.}
\begin{ruledtabular}
\begin{footnotesize}
\begin{tabular}{crrrrr}
$E_{\gamma}$ &                         $A_0$ &                            $a_1$ &                            $a_2$ &                            $a_3$ &                            $a_4$ \\
       (MeV) &                       (mb/sr) &                                  &                                  &                        ($=-a_1$) &                    ($=-(1+a_2)$) \\
\hline
        24.6 & 0.105 $\pm$ 0.007 $\pm$ 0.012 &    0.041 $\pm$ 0.108 $\pm$ 0.225 & $-$0.936 $\pm$ 0.134 $\pm$ 0.331 & $-$0.041 $\pm$ 0.108 $\pm$ 0.225 & $-$0.064 $\pm$ 0.134 $\pm$ 0.331 \\
        26.7 & 0.129 $\pm$ 0.003 $\pm$ 0.010 & $-$0.311 $\pm$ 0.032 $\pm$ 0.103 & $-$0.884 $\pm$ 0.044 $\pm$ 0.123 &    0.311 $\pm$ 0.032 $\pm$ 0.103 & $-$0.116 $\pm$ 0.044 $\pm$ 0.123 \\
        28.9 & 0.111 $\pm$ 0.002 $\pm$ 0.007 & $-$0.229 $\pm$ 0.027 $\pm$ 0.079 & $-$0.830 $\pm$ 0.037 $\pm$ 0.095 &    0.229 $\pm$ 0.027 $\pm$ 0.079 & $-$0.170 $\pm$ 0.037 $\pm$ 0.095 \\
        31.1 & 0.086 $\pm$ 0.001 $\pm$ 0.006 & $-$0.200 $\pm$ 0.026 $\pm$ 0.088 & $-$0.762 $\pm$ 0.036 $\pm$ 0.054 &    0.200 $\pm$ 0.026 $\pm$ 0.088 & $-$0.238 $\pm$ 0.036 $\pm$ 0.054 \\
        34.6 & 0.062 $\pm$ 0.002 $\pm$ 0.003 & $-$0.155 $\pm$ 0.037 $\pm$ 0.061 & $-$0.992 $\pm$ 0.050 $\pm$ 0.085 &    0.155 $\pm$ 0.037 $\pm$ 0.061 & $-$0.008 $\pm$ 0.050 $\pm$ 0.085 \\
        36.4 & 0.058 $\pm$ 0.001 $\pm$ 0.003 & $-$0.062 $\pm$ 0.034 $\pm$ 0.062 & $-$0.864 $\pm$ 0.047 $\pm$ 0.081 &    0.062 $\pm$ 0.034 $\pm$ 0.062 & $-$0.136 $\pm$ 0.047 $\pm$ 0.081 \\
        38.8 & 0.051 $\pm$ 0.001 $\pm$ 0.002 & $-$0.115 $\pm$ 0.042 $\pm$ 0.063 & $-$0.987 $\pm$ 0.059 $\pm$ 0.088 &    0.115 $\pm$ 0.042 $\pm$ 0.063 & $-$0.013 $\pm$ 0.059 $\pm$ 0.088 \\
        40.7 & 0.043 $\pm$ 0.001 $\pm$ 0.002 & $-$0.048 $\pm$ 0.044 $\pm$ 0.081 & $-$0.902 $\pm$ 0.061 $\pm$ 0.073 &    0.048 $\pm$ 0.044 $\pm$ 0.081 & $-$0.098 $\pm$ 0.061 $\pm$ 0.073 \\
        51.4 & 0.025 $\pm$ 0.002 $\pm$ 0.001 &    0.097 $\pm$ 0.097 $\pm$ 0.067 & $-$0.854 $\pm$ 0.138 $\pm$ 0.073 & $-$0.097 $\pm$ 0.097 $\pm$ 0.067 & $-$0.146 $\pm$ 0.138 $\pm$ 0.073 \\
        53.6 & 0.023 $\pm$ 0.001 $\pm$ 0.001 &    0.130 $\pm$ 0.087 $\pm$ 0.075 & $-$0.794 $\pm$ 0.121 $\pm$ 0.064 & $-$0.130 $\pm$ 0.087 $\pm$ 0.075 & $-$0.206 $\pm$ 0.121 $\pm$ 0.064 \\
        56.0 & 0.023 $\pm$ 0.001 $\pm$ 0.001 &    0.176 $\pm$ 0.089 $\pm$ 0.064 & $-$0.914 $\pm$ 0.128 $\pm$ 0.064 & $-$0.176 $\pm$ 0.089 $\pm$ 0.064 & $-$0.086 $\pm$ 0.128 $\pm$ 0.064 \\
        58.4 & 0.020 $\pm$ 0.001 $\pm$ 0.001 &    0.227 $\pm$ 0.102 $\pm$ 0.083 & $-$0.856 $\pm$ 0.139 $\pm$ 0.069 & $-$0.227 $\pm$ 0.102 $\pm$ 0.083 & $-$0.144 $\pm$ 0.139 $\pm$ 0.069 \\
        63.6 & 0.014 $\pm$ 0.001 $\pm$ 0.001 &    0.070 $\pm$ 0.151 $\pm$ 0.077 & $-$0.687 $\pm$ 0.186 $\pm$ 0.059 & $-$0.070 $\pm$ 0.151 $\pm$ 0.077 & $-$0.313 $\pm$ 0.186 $\pm$ 0.059 \\
        68.5 & 0.013 $\pm$ 0.001 $\pm$ 0.001 &    0.422 $\pm$ 0.103 $\pm$ 0.086 & $-$1.042 $\pm$ 0.135 $\pm$ 0.073 & $-$0.422 $\pm$ 0.103 $\pm$ 0.086 &    0.042 $\pm$ 0.135 $\pm$ 0.073 \\
\end{tabular}
\end{footnotesize}
\end{ruledtabular}
\end{table*}

\subsection{\label{subsection:angle_integrated_cross_section_data}Angle-integrated cross-section data}

A summary of the angle-integrated cross-section data is presented in Table
\ref{table:total}.

\begin{table}
\caption{\label{table:total}
A summary of the angle-integrated cross-section data.  The first uncertainty is
statistical and the second uncertainty is systematic.  See also 
Figure \ref{figure:angle_integrated_cross_section}.}
\begin{ruledtabular}
\begin{tabular}{cc}
$E_{\gamma}$ &                      $\sigma$ \\
       (MeV) &                          (mb) \\
\hline
        24.6 & 1.310 $\pm$ 0.134 $\pm$ 0.289 \\
        26.7 & 1.610 $\pm$ 0.064 $\pm$ 0.287 \\
        28.8 & 1.397 $\pm$ 0.049 $\pm$ 0.198 \\
        31.1 & 1.072 $\pm$ 0.038 $\pm$ 0.131 \\
        34.6 & 0.786 $\pm$ 0.034 $\pm$ 0.092 \\
        36.4 & 0.729 $\pm$ 0.035 $\pm$ 0.085 \\
        38.8 & 0.635 $\pm$ 0.033 $\pm$ 0.075 \\
        40.6 & 0.542 $\pm$ 0.031 $\pm$ 0.064 \\

        51.4 & 0.314 $\pm$ 0.046 $\pm$ 0.003 \\
        53.6 & 0.287 $\pm$ 0.039 $\pm$ 0.003 \\
        56.0 & 0.284 $\pm$ 0.038 $\pm$ 0.003 \\
        58.4 & 0.243 $\pm$ 0.037 $\pm$ 0.003 \\
        63.6 & 0.170 $\pm$ 0.036 $\pm$ 0.002 \\
        68.5 & 0.158 $\pm$ 0.019 $\pm$ 0.002 \\
\end{tabular}
\end{ruledtabular}
\end{table}

\bibliography{nilsson_etal}

\end{document}